# Uplink Macro Diversity of Limited Backhaul Cellular Network

Amichai Sanderovich*, Oren Somekh†, H. Vincent Poor†, and Shlomo Shamai (Shitz)*
* Department of Electrical Engineering, Technion, Haifa 32000, Israel
† Department of Electrical Engineering, Princeton University, Princeton, NJ 08544, USA
Email: amichi@tx.technion.ac.il, orens@princeton.edu, poor@princeton.edu, sshlomo@ee.technion.ac.il

**Abstract**

In this work new achievable rates are derived, for the uplink channel of a cellular network with joint multicell processing, where unlike previous results, the ideal backhaul network has finite capacity per-cell. Namely, the cell sites are linked to the central joint processor via lossless links with finite capacity. The cellular network is abstracted by symmetric models, which render analytical treatment plausible. For this idealistic model family, achievable rates are presented for cell-sites that use compress-and-forward schemes combined with local decoding, for both Gaussian and fading channels. The rates are given in closed form for the classical Wyner model and the soft-handover model. These rates are then demonstrated to be rather close to the optimal unlimited backhaul joint processing rates, already for modest backhaul capacities, supporting the potential gain offered by the joint multicell processing approach. Particular attention is also given to the low-SNR characterization of these rates through which the effect of the limited backhaul network is explicitly revealed. In addition, the rate at which the backhaul capacity should scale in order to maintain the original high-SNR characterization of an unlimited backhaul capacity system is found.

## I. Introduction

The growing demand for ubiquitous access to high-data-rate services has produced a significant amount of research analyzing the performance of wireless communications systems. Cellular systems are of major interest as the most common method for providing continuous services to mobile users, in both indoor and outdoor environments. In particular, the use of joint

This work was presented in part at the IEEE 2007 ISIT, June 2007, Nice, France.
This research was supported by a Marie Curie Outgoing International Fellowship and the NEWCOM++ network of excellence within the 6th and 7th European Community Framework Programmes, respectively, by the U.S. National Science Foundation under Grants ANI-03-38807 and CNS-06-25637, and the REMON consortium for wireless communication.

multicell processing (MCP) has been identified as a key tool for enhancing system performance (see [1] [2] and references therein for recent results on MCP).

Analysis of MCP has been so far based primarily on the assumption that all the base-stations (BSs) in the network are connected to a remote central processor (RCP) via an ideal backhaul network that is reliable, has infinite capacity and full connectivity. In this case, the set of BSs effectively acts as a multiantenna transmitter (downlink) or receiver (uplink) with the caveat that the antennas are geographically distributed over a large area. Since the assumption of an ideal backhaul network is quite unrealistic for large networks, more recently, there have been attempts to alleviate some of the conditions by considering alternative models. In [3] a limited-connectivity backhaul network model is studied in which only a subset of neighboring cells is connected to a central joint processor. In [4][5] a topological constraint is imposed in which there exist unlimited capacity links only between adjacent cells, and message passing techniques are employed in order to perform joint decoding in the uplink. Finally, [6] deals with practical aspects of limited capacity backhaul cellular systems incorporating MCP, where each BS quantizes its received signal and forwards it to the RCP via a finite-capacity reliable link. It is noted that [6] uses a simple quantization scheme that does not use the correlation between the received signals at neighboring BSs to reduce the compression rate.

Since its introduction in [7], the Wyner cellular model family has provided a framework for many studies dealing with multicell processing. Despite its simplicity, this model captures the essential structure of a cellular system and facilitates analytical treatment. The uplink channels of the Wyner linear and planar models are analyzed in [7] for optimal and linear minimum mean square error (MMSE) MCP receivers, and Gaussian channels. In [8], the Wyner model is extended to include fading channels and the performance of single and two cell-site processing under various situations is addressed. In [9] the results of [7] are extended to include flat fading channels (the reader is referred to [1] for a comprehensive survey of studies dealing with the Wyner model family).

As mentioned earlier, most works dealing with MCP assume that the backhaul network connecting the cell-sites to the RCP is error-free with infinite capacity. In this work we use the new, recently presented result from [10] to relax this assumption and allow each cell-site to connect to the RCP via a reliable error-free connection, but with limited capacity. Such a model suits cellular networks, where joint decoding can improve the overall network performance, with the underlying assumption that the received signals are forwarded to one location to be jointly processed. Since network resources are finite, in particular when the cell-sites are in fact “hot spots” with limited complexity, the inclusion of finite backhaul resources facilitates better prediction of the performance gain offered by MCP.

Recently, the common problem of nomadic terminals sending information to a remote destination via agents with lossless connections has been investigated in [11] and in [10] and then extended in [12] and [13] to include multiple-input multiple-output (MIMO) channels. The main difference between the scalar channels and the vector channels, is the inability to get a tight upper bound, due to the entropy power inequality. These works focus on the nomadic regime, in which the nomadic terminals use codebooks that are unknown to the agents, but are fully known to the remote destination. Such setting suits the uplink channel of the limited backhaul cellular system with MCP, where the oblivious cell-sites play the role of the agents.

Adhering to [10],[12], we assess the impact of limited backhaul capacity on the performance of two transmission schemes: (a) the “oblivious” scheme - in which the cell-sites are unaware of the users’ codebooks, and use a distributed Wyner-Ziv compress-and-forward scheme to send a compressed version of their received signal to the RCP for joint decoding; and (b) the “partial local decoding” scheme - in which the users split their messages into two parts, where the first part is decoded according to the “oblivious” scheme, while the second part is decoded locally by the relevant BSs (treated as a version of the broadcast channel).

Throughout this work we use two variants of the linear Wyner cellular setup [7], which provide a homogenous framework with respect to the mobile users and cell-sites. The first model is the circular Wyner model where the cells are arranged on a circle and each user “sees” three BSs, its local BS and the the two neighboring BSs. The second model is referred to as the circular soft-handoff (SH) model, in which the cells are arranged on a circle but the users “see” only two BSs, their local BS and the left neighboring BS. This setup focuses on users that are located on a cell edge, and thus are simultaneously received by two BSs. This type of situation is often referred to as soft-handoff. Due to its simplicity and the unique two-block diagonal structure of its channel transfer matrix, the SH model facilitates analytical treatment which provides additional insight. For both of these setups we are interested in the asymptotic scenario of infinitely many nodes, in which case the circular and linear models are equivalent. We consider both non-fading (Gaussian) and flat Rayleigh fading channels, although most of the results apply almost verbatim to other fading distributions.

Special attention is given to the low-signal to noise ratio (SNR) characterization of the resulting achievable rates. We apply the tools of [14] and provide closed-form expressions for the minimum energy per-bit required for reliable communication, and to the low-SNR rate slope for the suggested transmission schemes. In particular, the low-SNR parameters are expressed in a uniform way for both the limited and unlimited backhaul versions. This, in turn, reveals the effects of the limited backhaul network in a simple and concise manner. The high-SNR regime is also studied. In particular we provide the rate in which the backhaul capacity should scale with the SNR, in order for both schemes to preserve the high-SNR characterization of their parallel unlimited backhaul capacity counterparts.

The rest of the paper is organized as follow. In Section II we define the system models. Section III includes a short review of useful previous results that are used in the sequel. In Section IV we derive an achievable rate for oblivious cell-sites, while in Section V we extend the result to the case where cell-sites can perform local decoding. Numerical examples are presented in Section VI, which demonstrate the effect of limited backhaul capacity on MCP performance. Several proofs and the analysis of the low-SNR characterization, are relegated to the appendices.

## II. System model

Throughout this paper we consider two variants of the linear Wyner cellular uplink channel [7] (see also [15]): (a) the circular Wyner model, and (b) the “soft-handoff” model. Since both models resemble one another to a certain extent, the circular Wyner model is presented in full while the “soft-handoff” model is briefly described emphasizing its unique characterization. See Figure 1 for a simple illustration of the network topology.

### A. The Circular Wyner Model

This fully synchronized model includes $N$ identical cells, indexed by $j = 0, \dots, N-1$ arranged on a circle. Each cell includes $K$ identical single antenna mobile users, indexed $k = 1, \dots, K$, and a single-antenna base-station. According to the wide-band (WB) transmission scheme used, all transmitters simultaneously use all bandwidth. It is noted that the WB access protocol is optimal [8]. The users of the $j$th cell transmit $\{X_{j,k}\}_{k=1}^{K}$ into the channel. Each user sends the message $M_{j,k}$ to the RCP, where $M_{j,k} \in [1, \dots, 2^{nR_{j,k}}]$ and $R_{j,k}$ is defined as the communication rate. The rate region is defined to be achievable, if the probability of erroneous message in the RCP can be made arbitrary small, for sufficiently large block length.

Each cell-site receives the faded transmission of its cell's users with independently faded interference from the users of the adjacent cells and independent white additive noise. The received signal at the $j^{th}$ cell-site for time index $t$ reads

$$Y_j(t) = \sum_{k=1}^{K} a_{j,k}(t)\ X_j(t) + \alpha \sum_{k=1}^{K} \left( b_{j,k}(t)\ X_{[j-1]_N,k}(t) + c_{j,k}(t)\ X_{[j+1]_N,k}(t) \right) + Z_j(t), \quad \text{(2-1)}$$

where $[j]_N \triangleq j \mod N$, and the fixed inter-cell interference factor is $\alpha \in [0,\ 1]$. Each additive noise $Z_j(t)$ sample is a zero mean circularly-symmetric complex Gaussian random variable with unit variance, i.e. $Z_j(t) \sim CN(0,1)$. The users use zero mean circularly-symmetric complex Gaussian codebooks with average power $P/K$, $X_{j,k} \sim CN(0, \frac{P}{K})$. The fading coefficients $\{a_{j,k}, b_{j,k}, c_{j,k}\}_{k=1}^{K}$ are independent and identically distributed (i.i.d.) among different users and and can be viewed for each user as ergodic independent processes with respect to the time index. All the users are unaware of their instantaneous fading coefficients and are not allowed to cooperate in any way.

Using vector representation, expression (2-1) can be rewritten as (with the time index dropped for the sake of brevity)

$$Y_{\mathcal{N}} = H X_{\mathcal{N}\mathcal{K}} + Z_{\mathcal{N}}, \quad \text{(2-2)}$$

where $\mathcal{N} = \{0, \dots, N-1\}$ and $\mathcal{K} = \{1, \dots, K\}$. Accordingly, $Y_{\mathcal{N}} = \{Y_0, \dots, Y_{N-1}\}$ is the $N \times 1$ received signal vector, $X_{\mathcal{N}\mathcal{K}} = \{X_{0,1}, X_{0,2} \dots, X_{N-1,K-1}, X_{N-1,K}\}$ is the $NK \times 1$ transmit vector $X_{\mathcal{N}\mathcal{K}} \sim \mathcal{CN}(0, P/K I_{NK})$[1], and $Z_{\mathcal{N}} = \{Z_0, \dots, Z_{N-1}\}$ is the $N \times 1$ noise vector $Z_{\mathcal{N}} \sim \mathcal{CN}(0, I_N)$. The matrix $H$ is the $N \times NK$ channel transfer matrix defined by

$$H = \begin{pmatrix} \mathbf{a}_0 & \alpha\mathbf{c}_0 & 0 & \cdots & 0 & \alpha\mathbf{b}_0 \\ \alpha\mathbf{b}_1 & \mathbf{a}_1 & \alpha\mathbf{c}_1 & 0 & \cdots & 0 \\ 0 & \alpha\mathbf{b}_2 & \mathbf{a}_2 & \alpha\mathbf{c}_2 & \ddots & \vdots \\ \vdots & 0 & \alpha\mathbf{b}_3 & \ddots & \ddots & 0 \\ 0 & \vdots & \ddots & \ddots & \mathbf{a}_{N-2} & \alpha\mathbf{c}_{N-2} \\ \alpha\mathbf{c}_{N-1} & 0 & \cdots & 0 & \alpha\mathbf{b}_{N-1} & \mathbf{a}_{N-1} \end{pmatrix}, \quad \text{(2-3)}$$

[1]An $M \times M$ identity matrix is denoted by $I_M$

where, $\mathbf{a}_j = \{a_{j,1}, \ldots, a_{j,K}\}$, $\mathbf{b}_j = \{b_{j,1}, \ldots, b_{j,K}\}$ and $\mathbf{c}_j = \{c_{j,1}, \ldots, c_{j,K}\}$ are $1 \times K$ row vectors denoting the fading coefficients, experienced by the $K$ users of the $j$th, $[j-1]_N$th, and $[j+1]_N$th cells, respectively, and received by the $j$th cell-site.

The above description relates to the WB protocol where all users transmit simultaneously. For an intra-cell time-division multiple-access (TDMA) protocol, only one user is active per-cell, transmitting $1/K$ of the time using the total cell transmit power $P$. So for TDMA protocol we set $K = 1$ in (2-2) and (2-3).

Each cell-site is connected to the RCP through a unidirectional lossless link, with bandwidth of $C_j$ bits per channel use. The RCP, which is aware of all the users' fading coefficients, jointly processes the signals and decodes the messages sent by all the users of the cellular system, where the code rate in bits-per-channel-use of the $k$th user of the $j$th cell is $R_{j,k}$.

### B. The"Soft-Handoff" model (SH)

According to a circular variant of this model, $N$ cells are arranged on a circle. Unlike the Wyner model, the users are located on the cell edges and each user is received by the two closest cell-sites. Hence, the channel transfer matrix of this model is given by

$$H^{\mathrm{sh}} = \begin{pmatrix} \boldsymbol{a}_0 & \mathbf{0} & \cdots & \mathbf{0} & \alpha\boldsymbol{b}_0 \\ \alpha\boldsymbol{b}_1 & \boldsymbol{a}_1 & \ddots & \ddots & \vdots \\ \mathbf{0} & \ddots & \ddots & \ddots & \vdots \\ \vdots & \ddots & \ddots & \ddots & \mathbf{0} \\ \mathbf{0} & \cdots & \mathbf{0} & \alpha\boldsymbol{b}_{N-1} & \boldsymbol{a}_{N-1} \end{pmatrix}, \tag{2-4}$$

where $\mathbf{a}_j = \{a_{j,1}, \ldots, a_{j,K}\}$, and $\mathbf{b}_j = \{b_{j,1}, \ldots, b_{j,K}\}$ are $1 \times K$ row vectors denoting the fading coefficients, experienced by the signals of the $K$ users of the $j$th, and $[j-1]_N$th cells, respectively, when received by the $j$th cell-site. As with the Wyner model $\alpha \in [0,\ 1]$ represents the inter-cell interference factor.

All other definitions and assumptions made for the Wyner model in the previous sections hold for the "soft-handoff" model as well.

## III. Preliminaries

In this section we review previous results derived for the Wyner and SH models with unlimited backhaul capacity ($C \to \infty$), which will be useful in the sequel. As mentioned earlier, the central receiver is aware of all the users' codebooks and channel state information (CSI), and the users are not allowed to cooperate. Accounting for the underlying assumptions, the overall channel is a Gaussian multiple access channel (MAC) with $KN$ single antenna users and a one, $N$ distributed antenna receiver. Assuming an optimal joint MCP the per-cell sum-rate capacity of the unlimited backhaul Wyner model is given by

$$R = \frac{1}{N} E_H \left\{ \log_2 \det \left( I + \frac{P}{K} H H^{\dagger} \right) \right\} , \tag{3-1}$$

while the respective rate of the SH model is achieved by replacing $H$ with $H^{\rm sh}$ in (3-1). Due to the unique symmetric power profile of the channel transfer matrices involved, the rate (3-1) is known analytically only for certain special cases, which are reviewed in the following subsections[2].

Extreme SNR behavior of various rates of interest are considered throughout this work. In the low-SNR regime, rates are approximated by an affine expression which is characterized by two parameters: $\frac{E_b}{N_0}_{\rm min}$ is the minimal energy which is required to reliably transmit one bit; and $S_0$ is the slope (at $\frac{E_b}{N_0}_{\rm min}$) of the rate as a function of the SNR [14]. An affine expression is also used to approximate the rates in the high-SNR regime. In this case the rate is characterized by the high-SNR power slope $S_\infty$ (or multiplexing gain), and the high-SNR power offset $\mathcal{L}_\infty$ [17].

### A. Gaussian Channels, No Fading (nf)

Let us start with non-fading channels (so $a_{i,j}, b_{i,j}, c_{i,j}$ are constantly unity) and further focus on the asymptotic case where $N \to \infty$.

*1) Wyner Model:* The per-cell sum-rate capacity supported by the Wyner model is given by [7]

$$R_{\rm nf} = \int_0^1 \log_2\left(1 + P\left(1 + 2\alpha\cos(2\pi\theta)\right)^2\right) d\theta \ . \tag{3-2}$$

Since the entries of $\frac{1}{K}HH^\dagger$ are independent of $K$ for a fixed $P$ and non-fading channels, this rate is achievable (not uniquely) by both intra-cell TDMA and WB protocols. The low-SNR regime of (3-2) is characterized by [1]

$$\frac{E_b}{N_0}_{\rm min}^{\rm nf} = \frac{\log 2}{1 + 2\alpha^2} \quad ; \quad S_0^{\rm nf} = \frac{2(1+2\alpha^2)^2}{1 + 12\alpha^2 + 6\alpha^4} \ . \tag{3-3}$$

*2) SH Model:* The per-cell sum-rate of the SH setup is given by [18][19] (see also [20])

$$R_{\rm nf}^{\rm sh} = \log_2\left(\frac{1 + (1+\alpha^2)P + \sqrt{1 + 2(1+\alpha^2)P + (1-\alpha^2)^2P^2}}{2}\right) \ . \tag{3-4}$$

As with the Wyner model, intra-cell TDMA and WB protocols are capacity achieving protocols (not uniquely) under a total cell power constraint $P$. The low-SNR regime of (3-4) is characterized by [18][19]

$$\frac{E_b}{N_0}_{\rm min}^{\rm sh-nf} = \frac{\log 2}{1 + \alpha^2} \quad ; \quad S_0^{\rm sh-nf} = \frac{2(1+\alpha^2)^2}{1 + 4\alpha^2 + \alpha^4} \ . \tag{3-5}$$

[2]The reader is referred to [16] for further details on the intricate analytical issues concerning the spectrum of large random Hermitian finite-band matrices.

### B. Rayleigh Flat Fading (rf) channels

Introducing Rayleigh flat fading channels, intra-cell TDMA is no longer optimal and the WB protocol is known to be the capacity achieving transmission scheme [9]. A tight (for large $K$) upper bound for the per-cell sum-rate of the WB protocol is given by [9]. This rate is equal to the rate of a single-user SISO non-fading link with an additional channel gain due to the multiple cell-sites, which is $1+2\alpha^2$ for the Wyner model and $1+\alpha^2$ for the soft handover model.

*1) Wyner Model:* For the Wyner model, the upper bound for the per-cell sum-rate of the WB protocol is given by

$$R_{\mathrm{rf-lk}} = \log_2\left(1+(1+2\alpha^2)P\right) \ . \tag{3-6}$$

On the other hand, the low-SNR exact characterization of the rate is given by

$$\frac{E_b}{N_0}^{\mathrm{rf}}_{\mathrm{min}} = \frac{\log 2}{1+2\alpha^2} \quad ; \quad S_0^{\mathrm{rf}} = \frac{2}{1+\frac{1}{K}} \ . \tag{3-7}$$

It is noted that upper and lower moment bounds on the intra-cell TDMA protocol per-cell sum-rate are reported in [9] for fading channels as well. Since, these bounds are involved and are tight only in the low-SNR region, which is already covered by (3-7), we will not use them in the sequel.

*2) Soft-Handoff Model:* Turning to Rayleigh flat fading channels, the per-cell rate of the intra-cell TDMA scheme is derived in [18][19] for the special case of $\alpha = 1$ (based on a remarkable result of [21] calculating the capacity of an equivalent two tap time varying inter-symbol interference (ISI) channel)

$$R^{\mathrm{sh}}_{\mathrm{tdma-rf}} = \frac{\int_1^\infty (\log(x))^2 e^{-\frac{x}{P}} dx}{\mathrm{Ei}\left(\frac{1}{P}\right) P \log 2} \ , \tag{3-8}$$

where $\mathrm{Ei}(x) = \int_x^\infty \frac{\exp(-t)}{t} dt$ is the exponential integral function. For the WB protocol we have that the per-cell sum-rate is tightly upper bounded (with increasing $K$) by

$$R^{\mathrm{sh}}_{\mathrm{ub-rf}} = \log_2\left(\frac{1+(1+\alpha^2)P+\sqrt{\left(1+(1+\alpha^2)P\right)^2-4\alpha^2P^2/K}}{2}\right) \ . \tag{3-9}$$

This result was proved in [21] for $K = 1$ (intra-cell TDMA protocol) and was extended for arbitrary $K$ in [22].

Taking $K \to \infty$, this rate becomes

$$R^{\mathrm{sh}}_{\mathrm{rf-lk}} = \log_2\left(1+(1+\alpha^2)P\right) \ . \tag{3-10}$$

Finally, the low-SNR regime of the rate achieved by the WB protocol in the SH model is characterized in [18][19] by

$$\frac{E_b}{N_0}^{\mathrm{sh-rf}}_{\mathrm{min}} = \frac{\log 2}{1+\alpha^2} \quad ; \quad S_0^{\mathrm{sh-rf}} = \frac{2}{1+\frac{1}{K}} \ . \tag{3-11}$$

### C. Upper Bound

The following “cut-set-like” bound applies to both setups

**Proposition III.1** *For the two multicell setups at hand with equal limited backhaul $C$, the per-cell sum-rate is upper bounded by*

$$R_{\text{ub}} = \min\{C,\ R\}\ , \tag{3-12}$$

*where $R$ is the rate supported by the respective unlimited setup with MCP.*

*Proof:* This result follows by considering a cut-set bound [23] for two cuts, the first is by separating the central processor from the BSs, while the second is by separating the BSs from the MSs. For the second cut, it is easily verified that the normalized mutual information is equal to the per-cell sum-rate of the respective unlimited setup. We refer to this bound as a “cut-set-like” bound since we also account for the assumption of no MSs cooperation in the MS-BS cut ■

It is emphasized that the cut-set upper bound is general and particular bounds for the two setups under various conditions of interest, are achieved by replacing $R$ in (3-12) with the respective rates, reported in Section III. Furthermore, it is easily verified that replacing $R$ with an upper bound (such as (3-9)), results in a valid upper bound for the per-cell sum-rate.

## IV. Oblivious Cell-Sites

In this section we consider cell-sites that are oblivious to the users' codebooks and cannot perform local decoding. Instead, each cell-site forwards a compressed version of $Y_j$, namely $U_j$, to the RCP, through the lossless link of bandwidth $C_j$. The RCP then receives the compressed $\{U_j\}$ and decodes the messages sent by all the users.

### A. Gaussian Channels

Using similar argumentation as in [7], it is easy to verify that an intra-cell TDMA protocol is optimal in terms of the achievable throughput, for the non-fading homogenous model considered.

We begin by stating the following achievable rate-region for the MAC.

**Proposition IV.1** An achievable rate region for a general $N$ user MAC with oblivious $N$ cell-sites, connected to the RCP by error-free limited capacity links having capacities $\{C_j\}$ is given by

$$\forall \mathcal{L} \subseteq \{0,\ldots,N-1\} : \sum_{t\in\mathcal{L}} R_t \le \min_{\mathcal{S}\subseteq\mathcal{N}} \left\{ \sum_{j\in\mathcal{S}} [C_j - r_j] + I(X_{\mathcal{L}}; U_{\mathcal{S}^C} | X_{\mathcal{L}^C}) \right\}\ , \tag{4-1}$$

where

$$P_{X_{\mathcal{N}},U_{\mathcal{N}},Y_{\mathcal{N}}}(x_{\mathcal{N}}, u_{\mathcal{N}}, y_{\mathcal{N}}) = \prod_{j=1}^{N} P_{X_j}(x_j) \prod_{j=1}^{N} P_{Y_j|X_{\mathcal{N}}}(y_j|x_{\mathcal{N}}) \prod_{j=1}^{N} P_{U_j|Y_j}(u_j|y_j)\ , \tag{4-2}$$

and $r_j = I(Y_j; U_j | X_{\mathcal{N}})$.

An outline of the proof, based on [10], appears in part I of the Appendix and is given for any channel matrix $H$, including a random ergodic channel. When the channel is not a fading channel, such as in Proposition IV.1, the proof from Appendix I is applied by taking $H$ to be a known constant (which is also ergodic stationary process).

For the Gaussian channel we use $\{X_j, U_j\}$ that are complex Gaussian and also the joint probability (4-2) is Gaussian. It is noted that the Gaussian statistics are used due to the simplicity and relevancy of the reported results, with no claim of optimality. In fact, a better signalling approach is already suggested in [10], with direct implications here. For the Gaussian channel, the mutual information included in (4-1) reduces to [10][12]

$$I(X_{\mathcal{L}}; U_{\mathcal{S}^C} | X_{\mathcal{L}^C}) = \log_2 \det(I + P\text{diag}(1 - 2^{-r_j})_{j \in \mathcal{S}^C} H_{\mathcal{S}^C \mathcal{L}} H^*_{\mathcal{S}^C \mathcal{L}}), \tag{4-3}$$

where $H_{\mathcal{S}^C \mathcal{L}}$ is the transfer matrix between the output vector $Y_{\mathcal{S}^C}$ and the input vector $X_{\mathcal{L}}$, and $r_j$ are positive parameters that are subjected to optimization over $0 \leq r_j \leq C_j$. Focusing on the setup at hand, where $H_{i,j}$ is zero for $N - 1 > |i - j| > 1$ for both Wyner and SH models, equation (4-1) becomes

$$\sum_{t \in \mathcal{L}} R_t \leq \min_{\mathcal{S} \subseteq [\mathcal{L}+1]_N \cup [\mathcal{L}-1]_N \cup \mathcal{L}} \sum_{j \in \mathcal{S}} [C_j - r_j] + I(X_{\mathcal{L}}; U_{\mathcal{S}^C} | X_{\mathcal{L}^C}),$$

where $[\mathcal{L} \pm 1]_N \triangleq \{j : j = (i \pm 1) \mod N,\ i \in \mathcal{L}\}$. Let us define $H_{\mathcal{S}} = H_{\mathcal{S}\mathcal{N}}$, which is the transfer matrix between $X_{\mathcal{N}}$ and $Y_{\mathcal{S}}$.

Hereafter, we limit our attention to the symmetric case of $C_i = C$ for all cell-sites, and $R_t = R$ for all users. By symmetry and concavity, this limits the optimal $r_j$ to be invariant with respect to $j$: $r_j = r$, and the sum-rate inequality ($\mathcal{L} = \{0, \ldots, N-1\}$) to be the dominant inequality in (4-1).

Consequently we get the following.

**Corollary IV.2** An achievable rate for both Wyner and SH models with equal capacity links $C$, equal rate users and oblivious cell-sites is given by

$$R_{obl} = \frac{1}{N} \max_{0 \leq r} \left\{ \min_{\mathcal{S} \subseteq \mathcal{N}} \left\{ |\mathcal{S}|[C - r] + \log_2 \det \left( I + P(1 - 2^{-r}) H_{\mathcal{S}^C} H^*_{\mathcal{S}^C} \right) \right\} \right\}. \tag{4-4}$$

This rate is achieved by complex Gaussian $\{U_j, X_j\}$.

Next, we need to calculate the logarithm of the determinant in (4-4). In the case where no inter-cell interference is present ($\alpha = 0$), it is easily verified that $H_{\mathcal{S}^C} H^*_{\mathcal{S}^C}$ is an $|\mathcal{S}^C|$ identity matrix, and in this case the rate equals the rate achieved by an equivalent single-user single-agent Gaussian channel [10], which is given by

$$R_{\text{obl-g}} = \log_2 \left( 1 + P \frac{1 - 2^{-C}}{1 + P 2^{-C}} \right) \ . \tag{4-5}$$

For $\alpha > 0$, we focus on the case where the number of cells $N$ is large. An achievable rate for this asymptotic scenario is given by the following proposition, which is one of the central results in this paper.

**Proposition IV.3** *An achievable rate for the circular models with equal limited capacities, oblivious cell-sites and an infinite number of cells ($N \to \infty$), is given by*

$$R_{\text{obl}} = F(r^*), \tag{4-6}$$

*where $r^*$ is the solution of*

$$F(r^*) = C - r^*, \tag{4-7}$$

*and*

$$F(r) \triangleq \lim_{N\to\infty} \frac{1}{N} \log_2 \det\left(I + (1-2^{-r})PHH^\dagger\right) . \tag{4-8}$$

Notice that when $C \to \infty$, then also $r^* \to \infty$, and (4-6) reduces to the per-cell sum-rate capacity with optimal joint processing and unlimited backhaul capacity [7]. For finite $C$, the implicit equation (4-7) is easily solved numerically, since $F(r)$ is monotonic for the symmetric models at hand.

The following lemma is required for the proof of Proposition IV.3. This lemma is proved in part II of the Appendix for ergodic fading channels, where taking $H$ to be a known constant is a special case.

**Lemma IV.4** *Any subset $\mathcal{S}$ such that $|\mathcal{S}| = f(N)$ ($f : \mathbb{R}_+ \mapsto \mathbb{R}_+$, $\lim_{N\to\infty} \frac{f(N)}{N} = \lambda$, $0 \le \lambda \le 1$), which minimizes equation (4-4), when $N \to \infty$, includes only consecutive indices (considering also modulo operation).*

Denote a subset which contains only consecutive indices by $\mathcal{S}^{(c)}$.

*Proof of Proposition IV.3 (outline):* First, note that by applying Szegö's theorem [7], on $\log_2 \det(I + P(1-2^{-r})H_{\mathcal{S}^{(c)}}H^*_{\mathcal{S}^{(c)}})$ when $|\mathcal{S}^{(c)}| \to \infty$, we get the following simple explicit expression

$$\lim_{|\mathcal{S}^{(c)}|\to\infty} \frac{1}{|\mathcal{S}^{(c)}|} \log_2 \det(I + P(1-2^{-r})H_{\mathcal{S}^{(c)}}H^*_{\mathcal{S}^{(c)}}) = F(r).$$

Let us define $s = |\mathcal{S}^{(c)}|$, so that

$$\log_2 \det(I + P(1-2^{-r})H_{\mathcal{S}^{(c)}}H^*_{\mathcal{S}^{(c)}}) = sF(r) + \epsilon(s), \tag{4-9}$$

where $\lim_{s\to\infty} \epsilon(s)/s = 0$.

Secondly, from Lemma IV.4, when $N \to \infty$, a minimum for equation (4-4) is within the subspace of subsets that contain only consecutive indices $\{\mathcal{S}^{(c)}\}$. Combining (4-9), when $N \to \infty$, equation (4-4) becomes

$$\begin{aligned} R_{obl} &= \lim_{N\to\infty} \left\{ \max_{0\le r} \left\{ \min_{0\le s\le N} \left\{ \frac{N-s}{N}[C-r] + \frac{s}{N}F(r) + \frac{\epsilon(s)}{N} \right\} \right\} \right\} \\ &= \max_{0\le r} \left\{ \min_{0\le\lambda\le 1} \{(1-\lambda)[C-r] + \lambda F(r)\} \right\} . \end{aligned} \tag{4-10}$$

Since $F(r)$ is monotonically increasing, (4-10) is maximized by $r^*$, which is defined by $F(r^*) = C - r^*$. ■

*1) Low-SNR Characterization:* Next we study the low-SNR characterization of the oblivious schemes. The analysis is general and the results are used for various channels of interest. It is noted that throughout this section we assume that the finite backhaul capacity $C$ is much larger than the resulting rates.

Focusing on the low-SNR regime, where $P \ll 1$, the per-cell sum-rate of Proposition IV.3 in [bits/sec/Hz] is well approximated by the first three terms of its Taylor series:

$$F(r^*) \approx \dot{F}P(1-2^{-r^*})\log_2 e + \frac{1}{2}\ddot{F}P^2(1-2^{-r^*})^2\log_2 e + o(P^2) \ , \tag{4-11}$$

where $\dot{F} \triangleq \left.\frac{dF(\infty)}{dP}\right|_{P=0}$ and $\ddot{F} \triangleq \left.\frac{d^2F(\infty)}{dP^2}\right|_{P=0}$, are the first and second derivative of the unlimited backhaul rate function in [nats/sec/Hz] (when $r^* = \infty$) with respect to the SNR $P$ at $P = 0$. Substituting (4-11) in (4-7) we get the following equation:

$$\dot{F}P(1-2^{-r^*}) + \frac{1}{2}\ddot{F}P^2(1-2^{-r^*})^2 = (C - r^*)\log_e 2 \ . \tag{4-12}$$

Observing (4-7) it is clear that for low-SNR, $F(r^*)$ is small, which means that $C - r^* \ll 1$. Hence, $2^{C-r^*}$ is well approximated by

$$2^{C-r^*} \approx 1 + (C-r^*)\log_e 2 + o\left((C-r^*)^2\right) \ . \tag{4-13}$$

Substituting (4-11) into (4-12) and some additional algebra we get the following quadratic equation for the rate (in nats/dimension) of Proposition IV.3 in the low-SNR regime

$$\frac{1}{2}\ddot{F}P^2 2^{-2C}\ R^2 - \left(\dot{F}P2^{-C} + \ddot{F}P^2(1-2^{-C})2^{-C} + 1\right)\ R + \dot{F}P(1-2^{-C}) + \frac{1}{2}\ddot{F}P^2(1-2^{-C})^2 = 0 \ . \tag{4-14}$$

Neglecting the $R^2$ term we have that

$$R \approx \frac{\dot{F}P(1-2^{-C}) + \frac{1}{2}\ddot{F}P^2(1-2^{-C})^2}{\dot{F}P2^{-C} + \ddot{F}P^2(1-2^{-C})2^{-C} + 1} \ . \tag{4-15}$$

Finally, by applying the definitions of the low-SNR parameters of [14] to (4-15) and some additional algebra, we get the following proposition.

**Proposition IV.5** *The low-SNR characterization of the channel with the oblivious scheme and limited backhaul capacity $C$ is given by*

$$\frac{\mathrm{E_b}}{\mathrm{N_0}}_{\min} = \widetilde{\frac{\mathrm{E_b}}{\mathrm{N_0}}}_{\min}\frac{1}{1-2^{-C}} \quad ; \quad S_0 = \widetilde{S}_0\frac{1}{1+\widetilde{S}_0\frac{2^{-C}}{1-2^{-C}}} \ , \tag{4-16}$$

*where $\widetilde{\frac{\mathrm{E_b}}{\mathrm{N_0}}}_{\min}$ and $\widetilde{S}_0$ are the minimum transmitted energy per bit required for reliable communication, and the low-SNR slope of the unlimited channel $F(\infty)$, respectively.*

It is easily verified that with increasing backhaul capacity, the low-SNR characterization of the limited channel (4-16) coincides with that of the unlimited channel. Examining (4-16), it can also be verified that by allocating at least $C \approx 3.2$ [bits/sec/Hz] to the backhaul network, the minimum energy required for reliable communication of the limited channel will not increase by more than $0.5$ [dB] when compared to that of the unlimited backhaul.

Proposition IV.5 is especially useful in cases where (4-7) can not be solved explicitly. Nevertheless, we can use the few cases where (4-7) can be explicitly solved (e.g. the single-antenna single-agent Gaussian channel (4-5)) in order to validate the result of (4-16). Indeed, extracting the low-SNR characterization of the latter can be achieved by applying the definitions of [14] directly to (4-5):

$$\frac{\mathrm{E_b}}{\mathrm{N_0}}_{\min} = \frac{\log_e 2}{1 - 2^{-C}} \quad ; \quad S_0 = 2\frac{1 - 2^{-C}}{1 + 2^{-C}} \ . \tag{4-17}$$

The same result can be also derived by substituting the low-SNR characterization of the single-user single-antenna Gaussian case,

$$\frac{\mathrm{E_b}}{\mathrm{N_0}}_{\min} = \log_e 2 \quad ; \quad S_0 = 2 \ , \tag{4-18}$$

into (4-16).

*2) High-SNR characterization:* Here we study the high-SNR characterization of the oblivious scheme. Similar to the previous subsection, the high-SNR analysis is general and the results are applicable for various channels of interest. Specifically, the case of a fading channel is covered. From (4-7) it is evident that for a fixed backhaul capacity $C$ and increasing SNR $P$, the rate converges to $C$. Hence, for fixed $C$ the rate is finite and the system loses its multiplexing gain (i.e. $S_\infty = 0$). The latter can be also concluded immediately from the upper bound (3-12). Thus in the sequel, we are interested in the rate at which the backhaul capacity $C$ should scale with $P$, in order for the system to maintain it original (unlimited backhaul capacity setup) high-SNR characterization.

Following [17], the achievable rate of the unrestricted-backhaul system ($F(\infty)$) can be well approximated by the following affine expression for high SNR:

$$F(\infty) \cong S_\infty(\log_2 P - \mathcal{L}_\infty) \ , \tag{4-19}$$

where $S_\infty$ and $\mathcal{L}_\infty$ are the high-SNR parameters from [17].

Observing equation (4-3), for high SNR purposes, we can make the following argument: When $P' \triangleq P(1 - 2^{-r^\star})$ is very large, an unlimited system with $P'$ has the same high SNR characteristics as a limited-backhaul system with $P$. So that for very high $P(1 - 2^{-r^\star})$ we have

$$F(r^\star) \cong F(\infty)|_{P'=P(1-2^{-r^\star})} \cong S_\infty(\log_2 P - \mathcal{L}_\infty + \log_2(1 - 2^{-r^\star})). \tag{4-20}$$

Additionally, since we want the rate of the backhaul-limited system (4-7) to scale the same way as (4-19), we require the following asymptotic equivalences to hold:

$$R_{obl} = F(r^\star) \ \cong \ S_\infty(\log_2 P - \mathcal{L}_\infty) \tag{4-21}$$

$$R_{obl} = C - r^\star \ \cong \ S_\infty(\log_2 P - \mathcal{L}_\infty). \tag{4-22}$$

Taking $r^* \to \infty$ as $P \to \infty$, such that $P(1-2^{-r^*}) \to P$, the right hand side of the asymptotic equivalence (4-20) can replace the left hand side of the asymptotic equivalence of (4-21). On the other hand, (4-22) requires that

$$C \cong S_\infty(\log_2 P - \mathcal{L}_\infty) + r^*. \tag{4-23}$$

Thus taking $C \cong S_\infty(\log_2 P - \mathcal{L}_\infty) + \Theta(P)$, where $\Theta(P) \to \infty$ as $P \to \infty$ suffices to achieve the high-SNR characterization of the unrestricted-backhaul network. The exact scaling of $r^*$ can be very slow, for example $\log_2 \log_2 P$. Nonetheless, the larger the gap between $C$ and $F$ becomes ($\Theta(P)$ is increasing faster with $P$), the faster the asymptotic equivalence in the high-SNR is achieved.

**Corollary IV.6** *In order to preserve the high-SNR characterization of the original unlimited backhaul capacity setup ($S_\infty$ and $\mathcal{L}_\infty$), it is sufficient for the backhaul capacity to scale with the SNR on the order of $C(P) = S_\infty \log_2 P + \Theta(P)$, where $\Theta(P) \to \infty$ as $P \to \infty$, at arbitrary rate.*

In the next stage, we write closed-form expressions for $F$, using known results for both the Wyner and the SH models.

*3) The Wyner Model - Gaussian Channels:* For the Wyner model, $F(r)$ can be easily derived using (3-2), to be

$$F_{\text{nf}}(r) = \int_0^1 \log_2(1 + P(1-2^{-r})(1+2\alpha\cos 2\pi\theta)^2)d\theta \ . \tag{4-24}$$

So that for the unlimited scenario, the achievable rate is indeed the joint cell-site capacity:

$$R_{\text{nf}-\text{obl}} = F_{nf}(\infty) = \int_0^1 \log_2(1 + P(1+2\alpha\cos 2\pi\theta)^2)d\theta.$$

Next, we consider the low-SNR regime for the Wyner model. The main result stated in Proposition IV.5 is that the low-SNR characterization in this case can be expressed by the low-SNR characterization of the same channel but with unlimited backhaul. Using this result and the low-SNR characterization of the non-fading unlimited Wyner model (3-3), we get the low-SNR characterization of the per-cell sum-rate of the Wyner uplink channel with limited backhaul capacity:

$$\frac{\text{E}_\text{b}}{\text{N}_0}_{\text{min}} = \frac{\log 2}{(1+2\alpha^2)(1-2^{-C})} \quad ; \quad S_0 = \frac{2(1+2\alpha^2)^2(1-2^{-C})}{1+12\alpha^2+6\alpha^4+(1-4\alpha^2+2\alpha^4)2^{-C}} \ . \tag{4-25}$$

From (4-25), we see that the deleterious effect of limited backhaul is an increase in the minimum energy per-bit required for reliable communication and a decrease in the rate's low-SNR slope. This effect clearly diminishes when $C$ increases.

*4) The SH Model - Gaussian Channels:* Following similar arguments to those made for the Wyner setup, and capitalizing on the fact that the rate of the unlimited model is given in an explicit closed-form expression [19], we have the following closed form expression for the achievable per-cell sum-rate of the uplink SH model with oblivious BSs, average transmit power $P$, and equal limited backhaul capacity links $C$:

$$R_{\mathrm{sh-obl}} = \log_2\left(\frac{1+(1+\alpha^2)P+2\alpha^2 2^{-C}P^2+\sqrt{1+2(1+\alpha^2)P+\left((1-\alpha^2)^2+4\alpha^2 2^{-C}\right)P^2}}{2(1+2^{-C}P)(1+\alpha^2 2^{-C}P)}\right) . \tag{4-26}$$

See Appendix III for the derivation.

Next we consider the achievable rate $R_{sh-obl}$ under several asymptotic scenarios. For either increasing $C$ or increasing $P$ while the other is kept fixed, the rate coincides with the cut-set bound ($\min\{R_{sh}, C\}$).

Applying the definitions of [14] directly to (4-26) we get that for fixed $C$ the low-SNR regime of $R_{\mathrm{sh-obl}}$ is characterized by

$$\frac{E_b}{N_0}_{\min} = \frac{\log_e 2}{(1+\alpha^2)(1-2^{-C})} \quad ; \quad S_0 = \frac{2(1+\alpha^2)^2(1-2^{-C})}{1+4\alpha^2+\alpha^4+(1+\alpha^4)2^{-C}} . \tag{4-27}$$

As with the Wyner model, we see that the deleterious effects of the limited backhaul is an increase in the minimum energy per-bit required for reliable communication with a corresponding decrease in the rate's low-SNR slope. These effects clearly diminish when $C$ increases. It is noted that the same result can be obtained by applying Proposition IV.5 and substituting the low-SNR parameters of the unlimited SH model (3-5) in (4-16).

## B. Fading Channels

Upon the introduction of flat fading, the intra-cell TDMA protocol is no longer optimal even for the unlimited backhaul model, and the WB protocol is the capacity-region-achieving scheme (see [9]).

**Proposition IV.7** *The per-cell achievable ergodic sum-rate of the WB protocol deployed in the infinite circular model with equal limited capacities and in the presence of fading is given by* (4-6) *where*

$$F(r^*) = \lim_{N\longrightarrow\infty} \max_{\substack{r_i:\mathbb{C}^{N\times NK}\rightarrow\mathbb{R}_+ \\ \text{s.t. } \mathrm{E}r_i(\boldsymbol{H}_N)=r^*}} E\left(\frac{1}{N}\log_2\left(\det\left(I_N+\frac{P}{K}\mathrm{diag}\left(1-2^{-r_i(\boldsymbol{H})}\right)_{i=1}^N \boldsymbol{H}_N\boldsymbol{H}_N^\dagger\right)\right)\right) . \tag{4-28}$$

*Proof:* The proof for the fading channel follows along the same lines as the proof for the Gaussian channel in Proposition IV.3, and is based on [13]. The main differences consist of an additional conditioning on $H$ in all the mutual information expressions in Proposition IV.1 (where the proof of Proposition IV.1 in Appendix I already accounts for the fading), in an additional expectation with respect to $H$ in (4-4), and since the auxiliary variable $U$ depends also on the channel, so does $r$ (so the expectation will be also over $r_i$), and with an update to Lemma IV.4, such that it will be suited to the fading channel (where the proof of Lemma IV.4 in Appendix II, already accounts for the fading). ■

For the sake of compactness and usability, we use a lower bound to $F(r^*)$ from equation (4-28), by considering $r$ to be a constant $r^*$, regardless of the instantaneous channel $H$. This gives

$$F(r^*) = \lim_{N\longrightarrow\infty} \mathrm{E}\left(\frac{1}{N}\log_2\left(\det\left(I_N + \frac{P}{K}\left(1-2^{-r^*}\right)\boldsymbol{H}_N\boldsymbol{H}_N^\dagger\right)\right)\right) \ . \tag{4-29}$$

Since the channel is ergodic, for a large number of users $K$, this bound is tight (see [13]), where already at $K = 2$ there are 6 received signals at each cell-site and a very small gap is expected.

Unfortunately, the sum-rate $F(r^*)$ (and even $F(\infty)$) is explicitly known or can be bounded only for a few special cases. In the sequel we use the results presented in Section III-B to assess the impact of limited backhaul in these special cases.

*1) Wyner Model:* We start with the case where the number of users $K$ per-cell is large while the total cell average transmit power $P$ is fixed. In this case we have that

$$F(r^*) = R_{\mathrm{rf-lk}}(P(1-2^{-r^*}))\ , \tag{4-30}$$

where $R_{\mathrm{rf-lk}}$ is given in (3-6). Solving the fixed point equation $F(r^*) = C - r^*$ for (4-30), we get an explicit expression for the rate

$$R_{\mathrm{obl-rf-lk}} = \log_2\left(1 + \frac{(1+2\alpha^2)P(1-2^{-C})}{1+(1+2\alpha^2)P2^{-C}}\right) \ . \tag{4-31}$$

Hence, the rate of the limited network equals the rate of the single user Gaussian channel (4-5) but with enhanced power $(2\alpha^2+1)P$. It is noted that replacing $F(r^*)$ with an upper bound that increases with $P$ and equals zero when $P = 0$, provides an upper bound to the rate. Examining (4-31) it is easily verified that the rate achieves the cut-set bound when either $C$ or $P$ increases while the other is fixed. It is further noted that the rate (4-31) is also a tight upper bound for any arbitrary number of users per-cell.

Turning now to the low-SNR region with an arbitrarily number of users per-cell, we apply the general results of Proposition IV.5 and substitute the low-SNR parameters of the unlimited Wyner setup (3-7) in (4-16) to get the following per-cell sum-rate characterization for the Wyner Model with the WB protocol, in the presence of Rayleigh fading:

$$\frac{\mathrm{E_b}}{\mathrm{N_0}_{\min}} = \frac{\log 2}{(1+2\alpha^2)(1-2^{-C})} \quad ; \quad S_0 = \frac{2(1-2^{-C})}{1+\frac{1}{K}+\left(1-\frac{1}{K}\right)2^{-C}} \ . \tag{4-32}$$

Here also, the deleterious effect of the limited backhaul is again manifested in an increase in the minimum energy per-bit required for reliable communication and a decrease in the rate's low-SNR slope.

To conclude this section we note that the extreme results obtained for large $K$ and low-SNR can be easily extended to include a general fading distribution and are omitted for the sake of conciseness. In addition, for the intra-cell TDMA protocol ($K = 1$) we can use the moment bounds of [9] to provide respective lower and upper bounds to the rate in the limited backhaul case. Since these bounds are tight only in the low-SNR regime, which is basically covered by Proposition IV.5, they are omitted as well.

*2) The Soft-Handoff Model:* We start by claiming that the per-cell sum-rate supported by the SH model and WB protocol in the presence of fading is given by (4-29) while replacing the Wyner channel transfer matrix with the SH matrix (2-4). Here, as with the Wyner model, closed form expressions for the unlimited backhaul case are known only for a few limited cases (see Section III-B.2).

We consider first the intra-cell TDMA protocol ($K = 1$). Using the remarkable result of [21] which calculates the capacity for time-variant two-tap ISI channel (3-8), we have the following per-cell sum-rate capacity for the infinite circular SH model with oblivious BSs, an intra-cell TDMA protocol ($K = 1$), Rayleigh fading channels, and $\alpha = 1$

$$F(r^*) = R^{\mathrm{sh}}_{\mathrm{tdma-rf}}(P(1-2^{-r^*})) = \frac{\int_1^\infty (\log_e(x))^2 e^{-\frac{x}{P(1-2^{-r^*})}} dx}{\mathrm{Ei}\left(\frac{1}{P(1-2^{-r^*})}\right) P(1-2^{-r^*}) \log_e 2} \, , \tag{4-33}$$

where $\mathrm{Ei}(x) = \int_x^\infty \frac{\exp(-t)}{t} dt$ is the exponential integral function. Unfortunately, we are able to calculate the rate itself only numerically by solving (4-6) with (4-33).

Turning to the WB protocol where all $K$ users are active simultaneously, we use the upper bound of (3-9) to state the following upper bound:

$$R^{\mathrm{ub}}_{\mathrm{obl-rf}} = \log_2\left(\frac{1+P(1+\alpha^2)+2P^2\alpha^2 2^{-C}/K + \sqrt{(1+P(1+\alpha^2))^2 - 4P^2\alpha^2(1-2^{-C})/K}}{2\left(1+P(1+\alpha^2)2^{-C} + P^2\alpha^2 2^{-2C}/K\right)}\right) . \tag{4-34}$$

See Appendix IV for the derivation.

Next we consider the upper bound $R^{\mathrm{ub}}_{\mathrm{obl-rf}}$ under several asymptotic scenarios. For $C \to \infty$ and fixed $P$, $R^{\mathrm{ub}}_{\mathrm{obl-rf}}$ coincides with the respective unlimited setup (3-9). On the other hand, for $P \to \infty$ and fixed $C$, the upper bound $R^{\mathrm{ub}}_{\mathrm{obl-rf}} \to C$, achieving the cut-set bound. In addition, it is easily verified that $C$ should scale like $\log_2 P$ for $R^{\mathrm{ub}}_{\mathrm{obl-rf}}$ to achieve the optimal multiplexing gain of $1$. Finally, for increasing number of users $K \gg 1$, fixed total-cell power $P$ and finite $C$, the upper bound $R^{\mathrm{ub}}_{\mathrm{obl-rf}}$ reduces to

$$R^{\mathrm{ub}}_{\mathrm{obl-rf}} \underset{K\to\infty}{\longrightarrow} \log_2\left(1 + P\frac{(1+\alpha^2)(1-2^{-C})}{1+P(1+\alpha^2)2^{-C}}\right) \, , \tag{4-35}$$

which equals the rate of a limited Gaussian single user SISO channel (see (4-5)) with enhanced power $P(1+\alpha^2)$. It is noted that this result can be derived directly by setting $F(r^*) = R^{\mathrm{sh}}_{\mathrm{rf-lk}}(P(1-2^{-r*}))$ and solving $F(r^*) = C - r$, where $R^{\mathrm{sh}}_{\mathrm{rf-lk}}$ is the asymptotic expression (with increasing $K$) given in (3-10). Therefore, it is concluded that the bound $R^{\mathrm{ub}}_{\mathrm{obl-rf}}$ is tight for $K \gg 1$.

To assess the impact of limited backhaul in the low-SNR regime for Rayleigh fading channels we apply Proposition IV.5 and substitute the low-SNR parameters of the unlimited soft-handoff, to obtain

$$\frac{E_b}{N_0}_{\min} = \frac{\log 2}{(1+\alpha^2)(1-2^{-C})} \quad ; \quad S_0 = \frac{2(1-2^{-C})}{1+\frac{1}{K}+2^{-C}\left(1-\frac{1}{K}\right)} \ . \tag{4-36}$$

## V. Cell-sites with decoding

In order to better utilize the backhaul bandwidth between the cell-sites and the RCP, we consider using local decoding at the cell-sites. In this case the cell-sites should be aware of the associated codebooks, and thus do not operate in the nomadic regime [10]. It is noted that in general, decoding decreases the noise uncertainty, thus increasing the efficiency of backhaul usage.

In this section we present an intuitive, simple scheme which provides an achievable rate accounting for this local processing. According to this scheme, each user employs rate splitting and divides its message into two parts: one that is decoded at the RCP and another which is decoded at the local cell-site. In this case the message that is intended for the RCP to decode, interferes with the local decoding of the relevant message at the cell-site. Let the power used for the former be $\beta P$ and the latter $(1-\beta)P$, where $0 \le \beta \le 1$.

There are two strategies for the cell-site to execute: to decode only its local user's message, or to decode also the interfering users' messages, emerging from the neighboring cells (see [8] Section III.D). The locally decoded information rate is denoted by $R_d(\beta)$.

Forwarding the decoded information through the lossless links reduces the bandwidth available for compression, so the achievable rate is $R_{sd}(C)$ ($sd$ stands for *separate decoding*) given by

$$R_{sd}(C) = \max_{\beta}\left\{F_\beta(r_d^*) + \hat{R}_d(\beta)\right\} , \tag{5-1}$$

where

$$\hat{R}_d(\beta) = \min\{R_d(\beta), C\} \ , \tag{5-2}$$

$r_d^*$ is the solution of

$$F_\beta(r_d^*) = C - \hat{R}_d(\beta) - r_d^*, \tag{5-3}$$

and using equation (3-2), $F_\beta(r) = R_{\rm nf}(\beta P(1-2^{-r}))$ for the Wyner model, or using equation (3-4) $F_\beta(r) = R_{\rm nf}^{\rm sh}(\beta P(1-2^{-r}))$ for the SH model.

For $\alpha = 0$ this scheme is optimal, since there is no inter-cell interference and each cell-site can decode messages at the same rate as the RCP can.

Note that the rate $R_{sd}(C)$ is not concave in $C$ in general, and thus time-sharing may improve the achievable rate, which leads to the following proposition ($ch$ stands for the *convex-hull*).

**Proposition V.1** *An achievable rate of the rate-splitting scheme deployed in the infinite circular Wyner model with limited equal capacities $C$, is given by*

$$R_{sdch,1} = \max_{\lambda, C_1, C_2 :\ \lambda C_1 + (1-\lambda)C_2 \le C} \left\{\lambda R_{sd}(C_1) + (1-\lambda) R_{sd}(C_2)\right\} . \tag{5-4}$$

In fact, numerical calculations reveal that a good strategy is to do time-sharing between the two extreme approaches: using decoding at the cell-sites, with no decoding at the RCP, and decoding only at the RCP (4-10), rather than simultaneously using the mixed approach of (5-1). Thus, defining $t = \hat{R}_d(0)$, the rate $R_{sdch,1}$ of (5-4) can be written as

$$R_{dec} = \max_{r \geq r^*} \left\{ t + (C - t) \frac{F(r) - t}{F(r) + r - t} \right\} , \tag{5-5}$$

where $F(r) = R_{\rm nf}(P(1 - 2^{-r}))$, using equation (3-2) for the Wyner model and $F(r) = R^{\rm sh}_{\rm nf}(P(1 - 2^{-r}))$, using equation (3-4) for the SH model. The value of $r^*$ is calculated by (4-7). A more detailed derivation is given in Appendix V.

It is expected that decoding at the cell-site will be beneficial when $\alpha$ is small (low inter-cell interference), or when $C$ is small, so that decoding before transmission saves bandwidth, which otherwise would have been wasted on noise quantization.

*3) Low SNR Characterization:* To derive the low-SNR characterization ($P \ll 1$) of (5-5), with general rates $F(r)$ and $t$, we use (4-11) and

$$t \approx \dot{t} P \log_2 e + \frac{1}{2} \ddot{t} P^2 \log_2 e + o(P^2) \ , \tag{5-6}$$

where $\dot{t} \triangleq \frac{dt}{dP}\big|_{P=0}$ and $\ddot{t} \triangleq \frac{d^2 t}{dP^2}\Big|_{P=0}$, are the first and second derivative of the rate function in [nats/sec/Hz] with respect to the SNR $P$ at $P = 0$.

We start by noting that $(C - t) \approx C$, and since $r^* \approx C$ and the maximization is over $r \geq r^*$, we have also that $F(r) - t + r \approx r$. Hence, (5-5) can be rewritten in the low-SNR regime as

$$R \approx t + C \max_{r \geq C} \frac{F(r) - t}{r} \ . \tag{5-7}$$

By substituting (5-6) and (4-11) into (5-7) we get

$$R \approx \left\{ \dot{t} P + \frac{1}{2} \ddot{t} P^2 + C \max_{r \geq C} \frac{P(\dot{F}(1 - 2^{-r}) - \dot{t}) + \frac{1}{2} P^2 (\ddot{F}(1 - 2^{-r})^2 - \ddot{t})}{r} \right\} \log_e 2 \ . \tag{5-8}$$

Neglecting $P^2$ terms and recalling that $\dot{F} \geq \dot{t}$, it is easily verified that the maximization of (5-8) is achieved at

$$r_m = \max \{C, \ \tilde{r}_m\} \ , \tag{5-9}$$

where $\tilde{r}_m$ is the unique solution to the following fixed point equation with respect to $r$ :

$$2^{-r}(1 + r \log_e 2) = 1 - \frac{\dot{t}}{\dot{F}} \ , \tag{5-10}$$

which can be rewritten explicitly as

$$2^{-r}(1 + r \log_e 2) = 1 - \frac{\widetilde{\frac{\rm E_b}{\rm N_0}}_{\rm min}}{\frac{\rm E_b}{\rm N_0}{}^{\rm d}_{\rm min}} \ , \tag{5-11}$$

where $\widetilde{\frac{\mathrm{E_b}}{\mathrm{N_0}}}_{\min}$ and $\frac{\mathrm{E_b}}{\mathrm{N_0}}^{\mathrm{d}}_{\min}$ are the minimum transmitted energy per-bit of the local decoding scheme and the unlimited setup, respectively.

Using $r_m$, the optimal time-sharing parameter $\lambda_o$ (see (V.2)) can be approximated in the low-SNR region as

$$\lambda_o \approx 1 - \frac{C}{r_m} \ . \tag{5-12}$$

Furthermore, using $r_m$ the fixed equation (V.1) can be approximated in the low-SNR region by

$$F(r) = C' - r \ , \tag{5-13}$$

where $C'$ is the effective backhaul capacity, which can be approximated in the low-SNR region as

$$C' = \frac{C - \lambda t}{1 - \lambda} \approx r_m \ . \tag{5-14}$$

Finally, by applying the definitions of the low-SNR parameters of [14] to second equation of (V.1), using Proposition IV.5, and some additional algebra, we get the following proposition.

**Proposition V.2** *The general low-SNR characterization of the channel with central and local decoding scheme, and limited backhaul capacity $C$ is given by*

$$\begin{aligned} \frac{\mathrm{E_b}}{\mathrm{N_0}}^{\mathrm{dec}}_{\min} &= \frac{1}{\lambda_o \left(\frac{\mathrm{E_b}}{\mathrm{N_0}}^{\mathrm{d}}_{\min}\right)^{-1} + (1-\lambda_o)\left(\frac{\mathrm{E_b}}{\mathrm{N_0}}^{\mathrm{obl}}_{\min}(r_m)\right)^{-1}} \\ S_0^{\mathrm{dec}} &= \frac{\left(\frac{\mathrm{E_b}}{\mathrm{N_0}}^{\mathrm{dec}}_{\min}\right)^{-2}}{\lambda_o \left(S_0^{\mathrm{d}}\right)^{-1}\left(\frac{\mathrm{E_b}}{\mathrm{N_0}}^{\mathrm{d}}_{\min}\right)^{-2} + (1-\lambda_o)\left(S_0^{\mathrm{obl}}(r_m)\right)^{-1}\left(\frac{\mathrm{E_b}}{\mathrm{N_0}}^{\mathrm{obl}}_{\min}(r_m)\right)^{-2}} \ , \end{aligned} \tag{5-15}$$

*where the superscript $(\cdot)^{\mathrm{obl}}(r_m)$ indicates the low-SNR parameters of the oblivious scheme (calculated using Proposition IV.5 with $r_m$ replacing $C$), and the notation $(\cdot)^{\mathrm{d}}$ indicates the low-SNR parameters of the local decoding rate.*

Here, we are able to express the low-SNR parameters of the hybrid decoding scheme as functions of the low-SNR parameters of the respective local decoding and the oblivious schemes. The latter can be further expressed in terms of the low-SNR parameters of the non-limited scheme according to the results of Proposition IV.5. Examining (5-15) it is observed that for $\lambda_o = 0$ (or $C \geq \tilde{r}_m$ in (5-9)) the low-SNR parameters coincide with those of the oblivious scheme (Proposition IV.5). Hence, allocating resources to local decoding in the low-SNR regime is beneficial when $C$ is below a certain threshold $\tilde{r}_m$ (see (5-10)).

*4) High-SNR Characterization:* Similarly to the oblivious scheme, for any fixed backhaul capacity $C$, the rate of the partial local decoding scheme loses its multiplexing gain in the high-SNR regime. This is easily concluded from the upper-bound (3-12). So we focus on the case in which $C$ is allowed to increase with SNR and inter-cell interference is present (i.e. $\alpha > 0$). From the time-sharing behavior of (5-5), and since it is easily observed that local decoding does not attain the high SNR parameters of the unrestricted joint processing, it is concluded that the high-SNR solution to the time sharing (5-5), is only oblivious processing, which means $r \cong r^*$. Hence, in this case no local decoding is performed and all resources are allocated to the oblivious scheme. Thus, in the high-SNR regime the two schemes are equivalent and Cor. IV.6 holds for the the partial local decoding scheme as well. It is noted that when *no* inter-cell interference is present (i.e. $\alpha = 0$), MCP is not beneficial and all resources can equivalently be devoted to local decoding at the BSs. In order for the corresponding rate to maintain the high-SNR characterization of the original unlimited backhaul capacity setup, we should set $C = \hat{R}_d(0)$. Hence, $C$ should scale as $S_\infty^d(\log_2 P - \mathcal{L}_\infty^d)$ (where $S_\infty^d$, and $\mathcal{L}_\infty^d$ are the high-SNR parameters of $\hat{R}_d(0)$).

## A. Gaussian Channels

*1) The Wyner Model:* For the Wyner model, local decoding can be performed in three ways: decoding all three messages that arrive at the destination; decoding only the strongest message, treating the rest as interference; and decoding only the signals from the adjacent cells. This gives the following rate for local decoding [8]:

$$R_d = \max\left\{\log_2\left(1+\frac{(1-\beta)P}{1+(\beta+2\alpha^2)P}\right), \min\left\{\frac{1}{2}\log_2\left(1+\frac{(1-\beta)2\alpha^2 P}{1+\beta(1+2\alpha^2)P}\right), \frac{1}{3}\log_2\left(1+\frac{(1+2\alpha^2)(1-\beta)P}{1+\beta(1+2\alpha^2)P}\right)\right\}\right\} . \tag{5-16}$$

Substituting (5-16) in (5-1) and (5-4) gives the achievable rate.

To consider the low-SNR regime we apply the results of Proposition V.2 which approximate the low-SNR parameters of (5-5) for general rate expressions $F(r)$ and $\hat{R}_d(0)$. Noting that the low-SNR characterization of $\hat{R}_d(0)$ is given by

$$\frac{E_b}{N_0}^{\mathrm{d}}_{\min} = \log 2 \quad ; \quad S_0^{\mathrm{d}} = \frac{2}{1+4\alpha^2} , \tag{5-17}$$

the low-SNR characterization of (5-5) is obtained by substituting the low-SNR parameters of (4-25) and (5-17) in the general expressions (5-15), where $r_m = \max\{C, \tilde{r}_m\}$ and $\tilde{r}_m$ is the unique solution of

$$2^{-\tilde{r}_m}(1+\tilde{r}_m \log 2) = \frac{2\alpha^2}{1+2\alpha^2} . \tag{5-18}$$

Note that according to Proposition V.2 the time ratio dedicated to decoding at the BSs in the low-SNR region is $\lambda_{\mathrm{o}} = 1 - \frac{C}{r_m}$. In addition, examining (5-18) it is evident that $\tilde{r}_m$ is a decreasing function of the intra-cell interference factor $\alpha$. Therefore, it is concluded that in the low-SNR region, decoding also at the BSs is beneficial if the backhaul capacity $C$ is below a certain threshold which decreases with $\alpha$. For example, when there is no inter-cell interference $\alpha = 0$ then $\tilde{r}_m = \infty$ and decoding *only* at the BSs is optimal for any $C$. On the other hand, for $\alpha = 0.2$ numerical calculation reveals that $\tilde{r}_m \approx 2.15$ [bits]. Hence, incorporating decoding also at the BSs is beneficial when $C \lesssim 2.15$ [bits].

*2) The Soft-Handoff Model:* Similarly to the local decoding scheme applied for the Wyner model, here each user employs rate splitting and divides its message into two parts: one that is decoded at the RCP with power $(1-\beta)P$, and another that is decoded at the local cell-site with power $\beta P$. As before there are two strategies for the cell-site to execute: to decode only its local user's message; or to decode also the interfering users' messages emerging from the left neighboring cell (see [8] Section III.D). Such approach allows decoding of messages with rate

$$R_{\mathrm{d}}^{\mathrm{sh}} \quad = \quad \max\left\{ \log_2\left(1+\frac{(1-\beta)P}{1+(\beta+\alpha^2)P}\right), \frac{1}{2}\log_2\left(1+\frac{(1-\beta)(1+\alpha^2)P}{1+\beta(1+\alpha^2)P}\right)\right\} \quad . \tag{5-19}$$

Repeating steps (5-1)-(5-3) with the proper rate expressions for the SH model (expressions (5-19) and (4-26)) we get an achievable rate similar to (5-4).

As with the Wyner model, by using time sharing between the extreme cases $\beta = 0$ and $\beta = 1$, we get an explicit achievable rate expression $R_{\mathrm{dec}}^{\mathrm{sh}}$ similar to (5-5) with $t = \min\{C, R_{\mathrm{d}}^{\mathrm{sh}}(0)\}$. It is noted that unlike the Wyner model, here $r^*$ is explicitly given by $r^* = C - R_{\mathrm{obl}}^{\mathrm{sh}}$, where $R_{\mathrm{obl}}^{\mathrm{sh}}$ is given by (4-26).

To consider the low-SNR regime we apply the results of Proposition V.2. Noting that the low-SNR characterization of $\hat{R}_{\mathrm{d}}^{\mathrm{sh}}(0)$ is given by

$$\frac{E_b}{N_0}^{\mathrm{sh-d}}_{\mathrm{min}} = \log 2 \quad ; \quad S_0^{\mathrm{sh-d}} = \frac{2}{1+2\alpha^2} \ , \tag{5-20}$$

we obtain a similar result as with the Wyner model, with $r_m = \max\{C, \tilde{r}_m\}$ and where $\tilde{r}_m$ is now the unique solution of

$$2^{-\tilde{r}_m}(1+\tilde{r}_m\log 2) = \frac{\alpha^2}{1+\alpha^2} \ . \tag{5-21}$$

Similar observations as those made for the non-fading Wyner model are evident.

### *B. Fading Channels*

*1) The Wyner Model:* Introducing fading, and adhering to the simple scheme introduced in the previous section, decoding at an arbitrary BS (the cell index is omitted) yields the following

rate

$$R_{\mathrm{d-rf}} = \max\left\{ E\left(\log_2\left(1+\frac{(1-\beta)\frac{1}{K}\left|\boldsymbol{a}\right|^2 P}{1+\left(\beta\frac{1}{K}\left|\boldsymbol{a}\right|^2+\alpha^2(\frac{1}{K}\left|\boldsymbol{b}\right|^2+\frac{1}{K}\left|\boldsymbol{c}\right|^2)\right)P}\right)\right),\right.$$
$$\min\left\{\frac{1}{2}E\left(\log_2\left(1+\frac{(1-\beta)\alpha^2\left(\frac{1}{K}\left|\boldsymbol{b}\right|^2+\frac{1}{K}\left|\boldsymbol{c}\right|^2\right)P}{1+\beta\left(\frac{1}{K}\left|\boldsymbol{a}\right|^2+\alpha^2(\frac{1}{K}\left|\boldsymbol{b}\right|^2+\frac{1}{K}\left|\boldsymbol{c}\right|^2)\right)P}\right)\right),\right.$$
$$\left.\left.\frac{1}{3}E\left(\log_2\left(1+\frac{(1-\beta)\left(\frac{1}{K}\left|\boldsymbol{a}\right|^2+\alpha^2\left(\frac{1}{K}\left|\boldsymbol{b}\right|^2+\frac{1}{K}\left|\boldsymbol{c}\right|^2\right)\right)P}{1+\beta\left(\frac{1}{K}\left|\boldsymbol{a}\right|^2+\alpha^2(\frac{1}{K}\left|\boldsymbol{b}\right|^2+\frac{1}{K}\left|\boldsymbol{c}\right|^2)\right)P}\right)\right)\right\}\right\}, \quad (5\text{-}22)$$

where the expectations are taken with respect to the fading coefficient vectors $\boldsymbol{a}$, $\boldsymbol{b}$, and $\boldsymbol{c}$. [3]

Repeating steps (5-1)-(5-3) while setting $F(r)$ with (4-28) (or with (4-29) for a compact yet suboptimal rate), we can get a similar result as (5-4) for the fading channels with finite number of users per-cell $K$.

Focusing on the scenario of a large number of users per-cell $K \gg 1$ with a fixed total cell SNR $P$, it can be verified using the strong law of large numbers (SLLN) (see [8]) that (5-22) reduces to (5-16), and that repeating steps (5-1)-(5-3) while setting $F(r) = R_{\mathrm{rf-lk}}(P(1-2^{-r}))$, we get the following achievable rate:

$$R_{\mathrm{rf-lk-ld}} = \max_{\beta}\left\{\log_2\left(1+\frac{(1+2\alpha^2)\beta P(1-2^{-(C-\hat{R}_d(\beta))})}{1+(1+2\alpha^2)\beta P 2^{-(C-\hat{R}_d(\beta))}}\right)+\hat{R}_d(\beta)\right\}. \quad (5\text{-}23)$$

Moreover, using time sharing between the two extreme $\beta = 0$ and $\beta = 1$, we obtain

$$R_{rf-lk-ld2} = \max_{r\geq r^*}\left\{t+(C-t)\frac{\log_2\left(1+(1+2\alpha^2)P(1-2^{-r})\right)-t}{\log_2\left(1+(1+2\alpha^2)P(1-2^{-r})\right)+r-t}\right\}, \quad (5\text{-}24)$$

where (via (4-31))

$$r^* = \log_2\frac{2^C+(1+2\alpha^2)P}{1+(1+2\alpha^2)P}. \quad (5\text{-}25)$$

For the low-SNR regime, with a finite number of users per-cell ($K$ finite) we apply the results of Proposition V.2. The low-SNR characterization of $\hat{R}_{\mathrm{d-rf}}(0)$ is given by (see [1])

$$\frac{E_b}{N_0}^{\mathrm{rf-d}}_{\min} = \log 2 \quad ; \quad S_0^{\mathrm{rf-d}} = \frac{2}{1+4\alpha^2+\frac{1}{2K}}, \quad (5\text{-}26)$$

while (4-32) is then also used in the general expressions (5-15), where $r_m = \max\{C, \tilde{r}_m\}$ is equal to that of (5-18).

Similar conclusions as those of the non-fading channels are evident.

[3] It is noted that the expressions included in (5-22) can be rewritten as integrations over certain hypergeometric functions (see [8]). Since these integrals are numerically unstable, especially for large $K$ they are omitted here.

### C. The Soft-Handoff Model - Fading Channels

Introducing fading, and adhering to the simple scheme introduced in the previous section, decoding at an arbitrary BS yields the following rate

$$R^{\rm sh}_{\rm d-rf} = \max\left\{ E\left(\log_2\left(1+\frac{(1-\beta)\frac{1}{K}\left|\boldsymbol{a}\right|^2 P}{1+\left(\beta\frac{1}{K}\left|\boldsymbol{a}\right|^2+\alpha^2\frac{1}{K}\left|\boldsymbol{b}\right|^2\right)\right)P}\right)\right),\right.$$
$$\left.\frac{1}{2}E\left(\log_2\left(1+\frac{(1-\beta)\alpha^2\frac{1}{K}\left|\boldsymbol{b}\right|^2 P}{1+\beta\left(\frac{1}{K}\left|\boldsymbol{a}\right|^2+\alpha^2\frac{1}{K}\left|\boldsymbol{b}\right|^2\right)P}\right)\right)\right\}\ , \qquad (5\text{-}27)$$

where the expectations are taken with respect to the fading coefficient vectors $\boldsymbol{a}$ and $\boldsymbol{b}$.

As with the Wyner model, we repeat steps (5-1)-(5-3) setting $F(r)$ with specific expressions providing achievable rates for several cases of interest: (a) finite number of users $K$ - using (4-28) (or with (4-29) for a compact yet suboptimal rate) while replacing $H_N$ with the SH channel transfer matrix $H^{\rm sh}_N$; (b) upper bounds for finite number of users $K$ - using (4-34); and (c) TDMA with Rayleigh fading channels and $\alpha = 1$ - using the exact rate expression (4-33).

Focusing on the scenario of a large number of users per-cell $K \gg 1$ with fixed total cell SNR $P$, it is can be verified using the SLLN (see [8]) that (5-27) reduces to (5-19), and that by repeating steps (5-1)-(5-3) while setting $F(r) = R^{\rm sh}_{\rm rf-lk}(P(1-2^{-r}))$, we get similar expressions as those derived for the Wyner model while replacing the Wyner array power gain $(1+2\alpha^2)$ with the power gain of the SH array $(1+\alpha^2)$ in expressions (5-23), (5-24), and (5-25), respectively.

Finally, to consider the low-SNR regime we apply the results of Proposition V.2. Noticing that the low-SNR characterization of $\hat{R}^{\rm sh}_{\rm d-rf}(0)$ is given by

$$\left.\frac{E_b}{N_0}\right._{\min}^{\rm sh-rf-d} = \log 2 \quad ; \quad S_0^{\rm sh-rf-d} = \frac{2}{1+2\alpha^2+\frac{1}{K}}\ , \qquad (5\text{-}28)$$

we have that the low-SNR characterization is obtained by using the low-SNR characterization of the local decoding (5-28) and of the oblivious processing (4-36) in the general expression (5-15), where $r_m = \max\{C, \tilde{r}_m\}$ is equal to that of (5-21).

## VI. Numerical results

In this section we demonstrate the effects of limited-capacity backhaul links by several numerical examples. For the sake of conciseness, only the Wyner model is considered since similar conclusions apply for the SH model.

Achievable rates of the (a) oblivious scheme, (b) local-decoding scheme, and (c) unlimited setup, are plotted as functions of the inter-cell interference factor $\alpha$ for total cell power $P = 10$ [dB] and Gaussian (non-fading) channels, in Figures 2 and 3 for backhaul link capacity of $C = 3$ [bits/channel use] and $C = 6$, respectively. Upon examining the figures the deleterious effects of limited backhaul are revealed. In addition, the benefits of local decoding are evident for interference levels below a certain threshold which decreases with increasing values of $C$.

In particular, in Figure 2 where $C = 3$, the local decoding rate achieves the upper bound of the limited backhaul rate for interference levels below a certain threshold. This range reduces to a single point $\alpha = 0$ for large values of $C$ (e.g. Figure 3 where $C = 6$), where local decoding at the BSs alone is optimal since no inter-cell interference is present. On the other hand the oblivious approach cannot achieve the upper bound for finite values of $C$.

Introducing Rayleigh fading channels, the same rates are plotted for several values of the number of users per-cell $K$ in Figures 4 and 5 for backhaul link capacity of $C = 3$ [bits/channel use] and $C = 6$ respectively. Examining the figures, similar observations as those for the Gaussian channels are evident. As with the unlimited setup studied in [9], the rates in general increase with the number of users per-cell $K$, in the presence of fading.

In Figures 6 and 7, the upper-bound (which is the minimum between the unrestricted backhaul rate and the backhaul capacity) and the achievable rates of the oblivious and local-decoding schemes are plotted as functions of the total-cell power $P$ with $C = 6$ and $\alpha = 0.15$, for Gaussian and flat Rayleigh fading channels respectively. For Gaussian channels, the curves of the oblivious and the local decoding schemes are very close since the interference level is small, so decoding at the BSs brings marginal improvement. It is also observed that the rate of the oblivious scheme approaches the upper bound for low SNR values where $C$ is much larger than the unlimited rate, and for large SNR values where the unlimited rate is much larger than $C$. Similar behavior is observed for fading channels.

In Figures 8 and 9, the upper bound and the achievable rates are plotted as functions of the backhaul link capacity with $P = 10$ and $\alpha = 0.4$, for Gaussian and flat Rayleigh fading channels respectively. For both Gaussian and fading channels, it is observed that the rates of both schemes approach the upper bound for low values of $C$, where the unlimited rates are much higher than $C$, or for large values of $C$, where $C$ is much higher than the unlimited rates. Moreover, the benefit of the local decoding scheme is evident where in fact its rate achieves the upper bound $C$, below a certain threshold. As before, similar observations are made for the fading channels.

To conclude this section we verify the low-SNR regime analytical results derived for the oblivious and local decoding schemes with some numerical results derived for the Wyner setup with Gaussian channels. In Figures 10-12 the spectral efficiencies of the per-cell sum-rates of the (a) unlimited setup with optimal MCP, (b) the local decoding scheme, and (c) the oblivious scheme, are plotted for $C = 2,\ 4$ and $6$ [bits] respectively. For the two schemes, both the exact rates and low-SNR approximations (based on Propositions IV.5 and V.2) are plotted. Examining the curves it is observed that while the approximated minimum $\frac{E_b}{N_0}$ fairly matches the numerical results, the approximated low-SNR slope is somewhat optimistic when compared to the exact curves. Moreover, the benefit of the local decoding scheme is evident when the backhaul capacity $C$ is small.

## VII. Concluding Remarks

In this paper we have considered symmetric cellular models with limited backhaul capacity. Simple and tractable achievable rates have been derived for the case of cell-sites that use signal processing alone, and also when combined with local decoding. Both schemes considered no network planning, so inter-cell interference dominates. Closed form expressions have been developed for both the classical Wyner model and the SH model. Additional explicit low-SNR approximations for the achievable rates have also been derived. The rate at which the backhaul capacity should scale with SNR, in order for the various schemes to maintain their original high-SNR characterization, have also been derived. All results in this paper have included analysis for both Gaussian and Rayleigh fading channels. Numerical calculations reveal that unlimited optimal joint processing performance can be closely approached with a rather limited backhaul capacity, and that additional local decoding is beneficial only for low inter-cell interference levels.

## APPENDIX I
## PROOF OUTLINE OF PROPOSITION IV.1

The proof of Proposition IV.1 is based on the proof of Theorem 3 from [13]. We need the following lemma.

**Lemma I.1** *Generalized Markov Lemma*
*Let*

$$P_{A_{\mathcal{S}},Y_{\mathcal{S}}|H}(a_{\mathcal{S}},y_{\mathcal{S}}|h)=P_{Y_{\mathcal{S}}|H}(y_{\mathcal{S}}|h)\prod_{j\in\mathcal{S}}P_{A_j|Y_j,H}(a_j|y_j,h)\ . \tag{I.1}$$

*Given randomly generated* $\boldsymbol{y}_{\mathcal{S}}$ *according to* $P_{Y_{\mathcal{S}}|H}$*, for every* $j\in\mathcal{S}$*, randomly and independently generate* $N_j\geq 2^{nI(A_j;Y_j|H)}$ *vectors* $\tilde{\boldsymbol{a}}_j$ *according to* $\prod_{t=1}^{n}P_{A_j|H}(\tilde{a}_j(t)|h(t))$*, and index them by* $\tilde{\boldsymbol{a}}_j^{(v)}$ *(*$1\leq v\leq N_j$*). Then there exist* $|\mathcal{S}|$ *functions* $v_j^*=\phi_j(\boldsymbol{y}_j,\tilde{\boldsymbol{a}}_j^{(1)},\ldots,\tilde{\boldsymbol{a}}_j^{(N_j)})$ *taking values in* $[1\ldots N_j]$*, such that for any* $\epsilon>0$ *and sufficiently large* $n$*,*

$$\Pr((\{\boldsymbol{a}_j^{(v_j^*)}\}_{j\in\mathcal{S}},\boldsymbol{y}_{\mathcal{S}})\in\mathbf{T}_{\epsilon}(\boldsymbol{h}))\geq 1-\epsilon. \tag{I.2}$$

*Proof:* Lemma I.1 is Lemma 3.4 (Generalized Markov Lemma) in [24]. ■

### A. Code construction:

For every channel realization $\boldsymbol{h}$, determine the maximizing $\pi$. Fix $\delta>0$ and then

I) For every user $k$, within the $j^{th}$ cell,
- Randomly choose $2^{nR_{j,k}}$ vectors $\boldsymbol{x}_{j,k}$, with probability $P_{\boldsymbol{X}_{j,k}}(\boldsymbol{x}_{j,k})=\prod_t P_{X_{j,k}}(x_{j,k}(t))$.
- Index these vectors by $M_{j,k}$ where $M_{j,k}\in[1,2^{nR_{j,k}}]$.

II) For the compressor at the cell-sites
For every cell-site $j$ and every channel realizations matrix $\boldsymbol{H}$
- Randomly generate $2^{n[\hat{R}_j-C_j]}$ vectors $\boldsymbol{u}_j$ of length $n$ according to $\prod_t P_{U_j|H}(u_j(t)|h(t))$.
- Repeat the last step for $s_j=1,\ldots,2^{nC_j}$, define the resulting set of $\boldsymbol{u}_j$ of each repetition by $S_{s_j}$.
- Index all the generated $\boldsymbol{u}_j$ with $z_j\in[1,2^{n\hat{R}_j}]$. We will interchangeably use the notation $S_{s_j}$ for the set of vectors $\boldsymbol{u}_j$ as well as for the set of the corresponding $z_j$.
- Notice that the mapping between the indices $z_j$ and the vectors $\boldsymbol{u}_j$ depends on $\boldsymbol{h}$. So we will write $\boldsymbol{u}_j(z_j,\boldsymbol{h})$ to denote $\boldsymbol{u}_j$ which is indexed by $z_j$ for some specific $\boldsymbol{h}$.

*B. Encoding:*

Let $M = (M_{j,k})_{j=1,k=1}^{N,K}$ be the joint messages to be sent. The transmitters then send the corresponding $(\boldsymbol{x}_{j,k})_{j=1,k=1}^{N,K}$ to the channel.

*C. Processing at the cell-sites:*

The $j^{th}$ cell-site chooses any of the $z_j$ such that

$$\left(\boldsymbol{u}_j(z_j, \boldsymbol{h}), \boldsymbol{y}_j\right) \in \mathbf{T}_\epsilon^j(\boldsymbol{h}) \ , \tag{I.3}$$

where

$$\mathbf{T}_\epsilon^j(\boldsymbol{h}) \triangleq \left\{ \boldsymbol{u}_j, \boldsymbol{y}_j : \begin{array}{r} \forall u \in \mathcal{U}_j, h \in \mathcal{H}: \ \frac{1}{n}\left|N(u,h|\boldsymbol{u}_j,\boldsymbol{h}) - P_{U_j|H}(u|h)N(h|\boldsymbol{h})\right| < \frac{\epsilon}{|\mathcal{U}_j|} \\ \forall y \in \mathcal{Y}_j, h \in \mathcal{H}: \ \frac{1}{n}\left|N(y,h|\boldsymbol{y}_j,\boldsymbol{h}) - P_{Y_j|H}(y|h)N(h|\boldsymbol{h})\right| < \frac{\epsilon}{|\mathcal{Y}_j|} \\ u \in \mathcal{U}_j, y \in \mathcal{Y}_j, h \in \mathcal{H}: \ \frac{1}{n}\left|N(u,y,h|\boldsymbol{u}_j,\boldsymbol{y}_j,\boldsymbol{h}) - P_{U_j,Y_j|H}(u,y|h)N(h|\boldsymbol{h})\right| < \frac{\epsilon}{|\mathcal{Y}_j||\mathcal{U}_j|} \end{array} \right\} . \tag{I.4}$$

The event where no such $z_j$ is found is defined as the error event $E_1$. After deciding on $z_j$ the cell-site transmits $s_j$, which fulfills $z_j \in S_{s_j}$, to the RCP through the lossless link.

*D. Decoding (at the RCP):*

The destination retrieves $s_\mathcal{N} \triangleq (s_0, \ldots, s_{N-1})$ from the lossless links.
It then finds the set of indices $\hat{z}_\mathcal{N} \triangleq \{\hat{z}_1, \ldots, \hat{z}_\mathcal{N}\}$ of the compressed vectors $\hat{\boldsymbol{u}}_\mathcal{N}$ and the messages $M$ which satisfy

$$\begin{cases} \left(\boldsymbol{x}_{\mathcal{N},\mathcal{K}}(M_{\mathcal{N},\mathcal{K}}), \hat{\boldsymbol{u}}_\mathcal{N}(\hat{z}_\mathcal{N}, \boldsymbol{h})\right) \in \mathbf{T}_\epsilon^3(\boldsymbol{h}) \\ \hat{z}_\mathcal{N} \in S_{s_0} \times \cdots \times S_{s_{N-1}} \end{cases} \tag{I.5}$$

where $\mathbf{T}_\epsilon^3$ is defined in the standard way, as (I.4). If there is no such $\hat{z}_\mathcal{N}, \hat{M}_{\mathcal{N},\mathcal{K}}$, or if there is more than one, the destination chooses one arbitrarily. Define error $E_2$ as the event where $\hat{M}_{\mathcal{N},\mathcal{K}} \neq M_{\mathcal{N},\mathcal{K}}$.
Correct decoding means that the destination decides $\hat{M}_{\mathcal{N},\mathcal{K}} = M_{\mathcal{N},\mathcal{K}}$. An achievable rate region $R_{\mathcal{N},\mathcal{K}}$ was defined as when the RCP receives the transmitted messages with an error probability which is made arbitrarily small for sufficiently large block length $n$.

*E. Error analysis*

The error probability is upper bounded by

$$\Pr\{\text{error}\} = \Pr\left(E_1 \cup E_2\right) \leq \Pr(E_1) + \Pr(E_2) \ , \tag{I.6}$$

where

I) $E_1$ is the event that no $\boldsymbol{u}_j(z_j, \boldsymbol{h})$ is jointly typical with $\boldsymbol{y}_j$, and
II) $E_2$ is the event that there is a decoding error $\hat{M}_{\mathcal{N},\mathcal{K}} \neq M_{\mathcal{N},\mathcal{K}}$.

Next, we will upper bound the probabilities of the individual error events by arbitrarily small $\epsilon$.

*1) $E_1$:* According to Lemma I.1, the probability $\Pr\{E_1\}$ can be made as small as desired, for $n$ sufficiently large, as long as

$$\hat{R}_j > I(U_j; Y_j|H). \tag{I.7}$$

*2) $E_2$:* Consider the case where $\hat{M}_{\mathcal{L},\mathcal{Z}} \neq M_{\mathcal{L},\mathcal{Z}}$ and $\hat{z}_{\mathcal{S}} \neq z_{\mathcal{S}}$, where $\mathcal{S}, \mathcal{L} \subseteq \mathcal{N}$ and $\mathcal{Z} \subseteq \mathcal{K}$. There are

$$2^{n[\sum_{j\in\mathcal{L},k\in\mathcal{Z}} R_{j,k} + \sum_{i\in\mathcal{S}}[\hat{R}_i - C_i]]}$$

such vectors, and the probability of $(\boldsymbol{x}_{\mathcal{N},\mathcal{K}}(\hat{M}_{\mathcal{N},\mathcal{K}}), \boldsymbol{u}_{\mathcal{N}}(\hat{z}_{\mathcal{N}}))$ to be jointly typical is upper bounded by [13]

$$2^{n[h(X_{\mathcal{N},\mathcal{K}},U_{\mathcal{N}}|H) - h(U_{\mathcal{S}^C},X_{\{\mathcal{N},\mathcal{K}\}^C}|H) - \sum_{i\in\mathcal{S}} h(U_i|H) - \sum_{j\in\mathcal{L},k\in\mathcal{Z}} h(X_{j,k}) + \epsilon]},$$

where $h$ functions here also as the differential entropy. Thus $\Pr\{E_2\}$ can be made arbitrarily small as long as the rate region $R_{\mathcal{N},\mathcal{K}}$ for all $\mathcal{L}, \mathcal{S} \subseteq \mathcal{N}$ and $\mathcal{Z} \subseteq \mathcal{K}$ satisfies

$$\begin{aligned}
\sum_{j\in\mathcal{L},k\in\mathcal{Z}} R_{j,k} &< \sum_{i\in\mathcal{S}}[C_i - \hat{R}_i + h(U_i|H) - h(U_i|H, X_{\mathcal{N},\mathcal{K}})] - h(X_{\mathcal{L},\mathcal{Z}}|X_{\{\mathcal{L},\mathcal{Z}\}^C}, U_{\mathcal{S}^C}, H) \\
&\quad + \sum_{j\in\mathcal{L},k\in\mathcal{Z}} h(X_{j,k}|X_{\{\mathcal{L},\mathcal{Z}\}^C}, H) \\
&= \sum_{i\in\mathcal{S}}[C_i - I(Y_i; U_i|X_{\mathcal{N},\mathcal{K}}, H)] + I(U_{\mathcal{S}^C}; X_{\mathcal{L},\mathcal{Z}}|X_{\{\mathcal{L},\mathcal{Z}\}^C}, H)\ ,
\end{aligned} \tag{I.8}$$

where (I.8) is due to of the following equalities:

$$\begin{aligned}
h(U_i|Y_i) &= h(U_i|Y_i, X_{\mathcal{N},\mathcal{K}})\ , \\
h(X_{\mathcal{L},\mathcal{Z}}) &= h(X_{\mathcal{L},\mathcal{Z}}|X_{\{\mathcal{L},\mathcal{Z}\}^C})\ . \\
h(X_{\mathcal{L},\mathcal{Z}}, U_{\mathcal{S}}|X_{\{\mathcal{L},\mathcal{Z}\}^C}, U_{\mathcal{S}^C}) &= h(X_{\mathcal{L},\mathcal{Z}}|X_{\{\mathcal{L},\mathcal{Z}\}^C}, U_{\mathcal{S}^C}) + \sum_{i\in\mathcal{S}} h(U_i|X_{\mathcal{N},\mathcal{K}})\ .
\end{aligned}$$

Equation (I.8) completes the proof. ■

## APPENDIX II
## PROOF OF LEMMA IV.4

We prove that at least one $\mathcal{S}$ which minimizes

$$\lim_{N\to\infty} \frac{1}{N} I(X_{\mathcal{N},\mathcal{K}}; U_{\mathcal{S}}) = \lim_{N\to\infty} \frac{1}{N} \log_2 \det(I + P' H_{\mathcal{S}} H_{\mathcal{S}}^*),$$

when $|\mathcal{S}| = f(N)$, ($f : \mathbb{R}_+ \mapsto \mathbb{R}_+$, $\lim_{N\to\infty} \frac{f(N)}{N} = \lambda$, $0 \leq \lambda \leq 1$), is composed of only consecutive indices. Following the method used in [25] to derive a lower bound on the capacity of the Gaussian erasure channel, the proof here uses an analogy between the multi-cell setup and an inter-symbol interference (ISI) channel, combined with a recently reported relationship between the MMSE and the mutual information [26].

*Proof:* Denote by $E_i$, the MMSE incurred when estimating $H_i X_{\mathcal{N},\mathcal{K}}$ from $U_{\mathcal{N}}, h$. Further denote by $E_i(\mathcal{S})$, the MMSE incurred when estimating $H_i X_{\mathcal{N},\mathcal{K}}$ from $U_{\mathcal{S}}, h$. Naturally

$$\forall h, \mathcal{S} \subseteq \mathcal{N},\ i \in \mathcal{S}:\ E_i(\mathcal{S}) \geq E_i\ , \tag{II.1}$$

and also

$$i \notin \mathcal{S}:\ E_i(\mathcal{S}) = 0\ . \tag{II.2}$$

Next, we use the following relationship between the MMSE and the mutual information [26], to write

$$\frac{d}{dP} I(X_{\mathcal{N}}; U_{\mathcal{S}}|H) = \sum_{i=0}^{N-1} E_i(\mathcal{S})\ . \tag{II.3}$$

From (II.1) and the ergodicity of the channel, we can write

$$\sum_{i=0}^{N-1} E_i(\mathcal{S}) \geq \frac{f(N)}{N} \sum_{i=0}^{N-1} E_i\ . \tag{II.4}$$

Combining (II.3) and (II.4) yields

$$I(X_{\mathcal{N}}; U_{\mathcal{S}}|H) \geq \frac{f(N)}{N} \int_0^{P'} \sum_{i=0}^{N-1} E_i dP = \frac{f(N)}{N} I(X_{\mathcal{N}}; U_{\mathcal{N}}|H). \tag{II.5}$$

On the other hand, in the asymptotic regime, for consecutive indices set $\mathcal{S}^{(c)}$, where $\lim_{N\to\infty} \frac{|\mathcal{S}^{(c)}|}{N} = \lambda$, we have

$$\lim_{N\to\infty} \frac{1}{N} \sum_{i \in \mathcal{S}^{(c)}} E_i(\mathcal{S}^{(c)}) = \lambda \lim_{N\to\infty} \frac{1}{N} \sum_{i=0}^{N-1} E_i. \tag{II.6}$$

This is because the equivalent ISI channel is stationary, and since the right hand side of (II.6) exists. By integrating both sides of equation (II.6) we get that

$$\lim_{N\to\infty} \frac{1}{N} I(X_{\mathcal{N}}; U_{\mathcal{S}^{(c)}}|H) = \lambda \lim_{N\to\infty} \frac{1}{N} I(X_{\mathcal{N}}; U_{\mathcal{N}}|H). \tag{II.7}$$

Equation (II.7) together with (II.5) proves the lemma. ■

## Appendix III
## Proof of the achievable rate of the SH model with limited backhaul and Gaussian channels, equation (4-26)

Using arguments similar to these used for the Wyner setup, the per-cell sum-rate of the limited soft-handoff setup is given by Proposition IV.3, with $F(r^*) = R_{\rm nf}^{\rm sh}(P(1-2^{-r^*})$ where $R_{\rm nf}^{\rm sh}$ is the rate of the unlimited setup given in (3-4). Due to the explicit simple form of $F(r^*)$, the fixed point equation (4-7) reduces to the following quadratic equation:

$$(1+P2^{-C})(1+\alpha^2 P 2^{-C})x^2 - \left(1 + (1+\alpha^2)P + 2\alpha^2 P^2 2^{-C}\right)x + \alpha^2 P^2 = 0 \tag{III.1}$$

where $x = 2^{C-r^*}$, and its roots are given by

$$x_{1,2} = \frac{1+(1+\alpha^2)P + 2\alpha^2 2^{-C}P^2 \pm \sqrt{1+2(1+\alpha^2)P + \left((1-\alpha^2)^2 + 4\alpha^2 2^{-C}\right)P^2}}{2(1+2^{-C}P)(1+\alpha^2 2^{-C}P)} \ . \tag{III.2}$$

For $x_1$ we have the following set of inequalities

$$\begin{aligned} x_1 &\geq \frac{1+(1+\alpha^2)2^{-C}P + 2\alpha^2 2^{-2C}P^2 + \sqrt{1+2(1+\alpha^2)2^{-C}P + \left((1-\alpha^2)^2 + 4\alpha^2\right)2^{-2C}P^2}}{2(1+2^{-C}P)(1+\alpha^2 2^{-C}P)} \\ &= \frac{1+(1+\alpha^2)2^{-C}P + 2\alpha^2 2^{-2C}P^2 + \sqrt{1+2(1+\alpha^2)2^{-C}P + (1+\alpha^2)^2 2^{-2C}P^2}}{2(1+2^{-C}P)(1+\alpha^2 2^{-C}P)} \\ &= \frac{1+(1+\alpha^2)2^{-C}P + 2\alpha^2 2^{-2C}P^2 + \sqrt{\left(1+(1+\alpha^2)2^{-C}P\right)^2}}{2(1+2^{-C}P)(1+\alpha^2 2^{-C}P)} \\ &= \frac{1+(1+\alpha^2)2^{-C}P + \alpha^2 2^{-2C}P^2}{(1+2^{-C}P)(1+\alpha^2 2^{-C}P)} = 1 \ . \end{aligned} \tag{III.3}$$

Hence, choosing $+$ in (III.2) yields a valid solution with $r^*$ which is also smaller than the backhaul capacity $C$ for all values of $P$, $C$ and $\alpha$ (i.e. $x_1 \geq 1, \ \forall P, \ C, \ \alpha$). Recalling that the rate equals $C - r^*$ (see (4-7)) completes the derivation.

## Appendix IV
## Proof of the achievable rate of the SH model with limited backhaul and fading channels, equation (4-34)

We start by observing that replacing $F(r^*)$ in (4-6) with an upper bound $F^{ub}(r) \geq F(r^*)$ where $F^{ub}(0) = 0$, provides a valid solution to (4-7), which is also an upper bound on $F(r^*)$. Setting $F(r^*) = R^{\rm sh}_{\rm ub-rf}(P(1-2^{-r^*}))$ where $R^{\rm sh}_{\rm rf-ub}$ is given in (3-9) and solving the fixed point equation (4-7), we get the following quadratic equation

$$\left(1+P(1+\alpha^2)2^{-C} + P^2\alpha^2 2^{-2C}/K\right)x^2 - \left(1+P(1+\alpha^2) + 2\alpha^2P^2 2^{-C}/K\right)x + P^2\alpha^2/K = 0 \ , \tag{IV.1}$$

where $x = 2^{C-r^*}$, and its roots are given by

$$x_{1,2} = \frac{1+P(1+\alpha^2) + 2P^2\alpha^2 2^{-C}/K \pm \sqrt{\left(1+P(1+\alpha^2)\right)^2 - 4P^2\alpha^2(1-2^{-C})/K}}{2\left(1+P(1+\alpha^2)2^{-C} + P^2\alpha^2 2^{-2C}/K\right)} \ . \tag{IV.2}$$

For $x_1$ we have the following set of inequalities

$$
\begin{aligned}
x_1 &\geq \frac{1+P(1+\alpha^2)+2P^2\alpha^2 2^{-C}/K \pm \sqrt{\left(1+P(1+\alpha^2)\right)^2 - 4P^2\alpha^2(1-2^{-C})}}{2\left(1+P(1+\alpha^2)2^{-C}+P^2\alpha^2 2^{-2C}/K\right)} \\
&= \frac{1+P(1+\alpha^2)+2P^2\alpha^2 2^{-C}/K + \sqrt{1+2P(1+\alpha^2)+P^2\left((1-\alpha^2)^2+4\alpha^2 2^{-C}\right)}}{2\left(1+P(1+\alpha^2)2^{-C}+P^2\alpha^2 2^{-2C}/K\right)} \\
&\geq \frac{1+P(1+\alpha^2)2^{-C}+2P^2\alpha^2 2^{-2C}/K + \sqrt{1+2P(1+\alpha^2)+P^2\left((1-\alpha^2)^2+4\alpha^2\right)2^{-C}}}{2\left(1+P(1+\alpha^2)2^{-C}+P^2\alpha^2 2^{-2C}/K\right)} \\
&\geq \frac{1+P(1+\alpha^2)2^{-C}+2P^2\alpha^2 2^{-2C}/K + \sqrt{1+2P(1+\alpha^2)2^{-C}+P^2(1+\alpha^2)^2 2^{-2C}}}{2\left(1+P(1+\alpha^2)2^{-C}+P^2\alpha^2 2^{-2C}/K\right)} \\
&= \frac{1+P(1+\alpha^2)2^{-C}+2P^2\alpha^2 2^{-2C}/K + \sqrt{\left(1+P(1+\alpha^2)2^{-C}\right)^2}}{2\left(1+P(1+\alpha^2)2^{-C}+P^2\alpha^2 2^{-2C}/K\right)} \\
&= \frac{2\left(1+P(1+\alpha^2)2^{-C}+P^2\alpha^2 2^{-2C}/K\right)}{2\left(1+P(1+\alpha^2)2^{-C}+P^2\alpha^2 2^{-2C}/K\right)} = 1\ .
\end{aligned}
\tag{IV.3}
$$

Hence, choosing $+$ in (IV.2) yields a valid solution with $r^*$ which is also smaller than the backhaul capacity $C$ for all values of $P$, $C$ and $\alpha$ (i.e. $x_1 \geq 1,\ \forall P,\ C,\ \alpha$). Recalling that the rate upper bound equals $C - r^*$ (see (4-7)) completes the derivation.

## Appendix V
## Derivation of Expression (5-5)

The derivation is based on time-sharing between the point $(t,t)$ and the concave curve $(F(r)+r, F(r))$ (using power $P$ in both techniques). The first point is achieved by using local decoding at the cell-site, while the second is achieved by using oblivious processing at the cell-sites and RCP decoding. Using only local decoding is optimal when $R_d(0) = C$. When $R_d(0) < C$, it is worthwhile to use time sharing with some point $(F(r') + r', F(r'))$. This means

$$
\begin{cases}
\lambda t + (1-\lambda)(F(r') + r') = C \\
\lambda t + (1-\lambda)F(r') = R_{dec}.
\end{cases}
\tag{V.1}
$$

From first equation of (V.1), we have

$$
\lambda = \frac{C - (F(r') + r')}{t - (F(r') + r')}\ , \tag{V.2}
$$

and assigning this value back to the second equation of (V.1) we obtain

$$
\begin{aligned}
R &= \frac{C-(F(r')+r')}{t-(F(r')+r')}t + \frac{t-C}{t-(F(r')+r')}F(r') \\
&= t - \frac{C-t}{t-(F(r')+r')}t + \frac{t-C}{t-(F(r')+r')}F(r') \\
&= t + (C-t)\frac{F(r')-t}{F(r')+r'-t}\ .
\end{aligned}
\tag{V.3}
$$

Next, we would like to optimize over $r'$, such that we get the maximal rate. Considering that $0 \leq \lambda \leq 1$, $r'$ must be larger than $r^*$, thus limiting the optimization range.

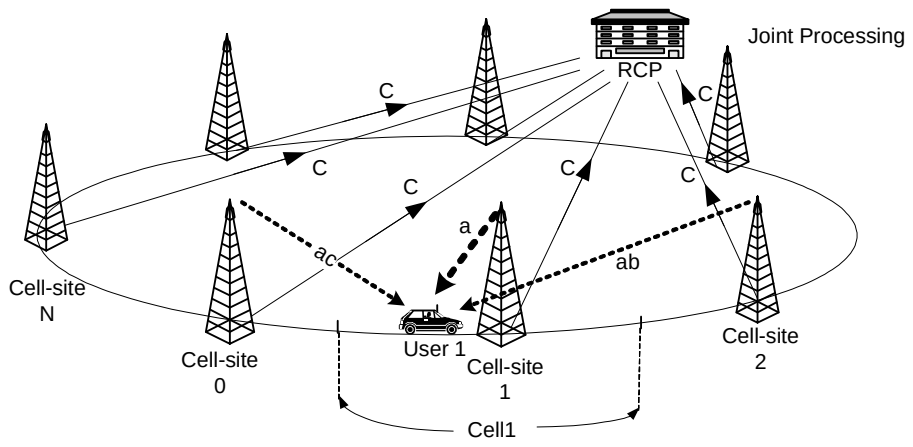

Fig. 1. Simple finite circular symmetric cellular model, with limited finite capacity to a central processing unit. One user is also drawn to demonstrate the model.

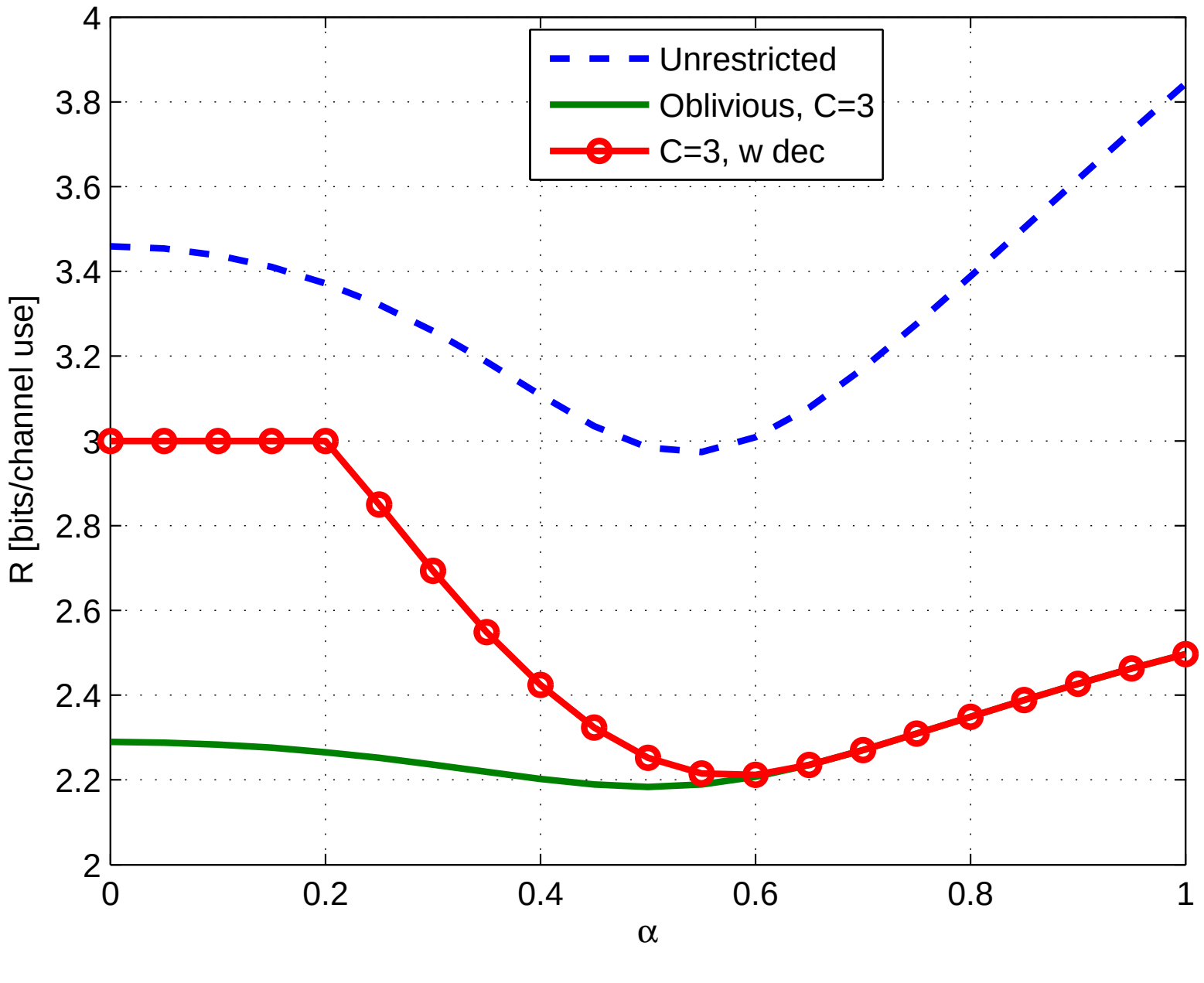

Fig. 2. The achievable rate for limited capacity $C = 3$ bits per channel use and oblivious processing (solid line) or combined with local decoding (circle marker), as compared with the unrestricted backhaul rate (dashed line), as a function of the interference level $\alpha$. The signal to noise ratio is $P = 10$ dB, and the channel is additive white Gaussian.

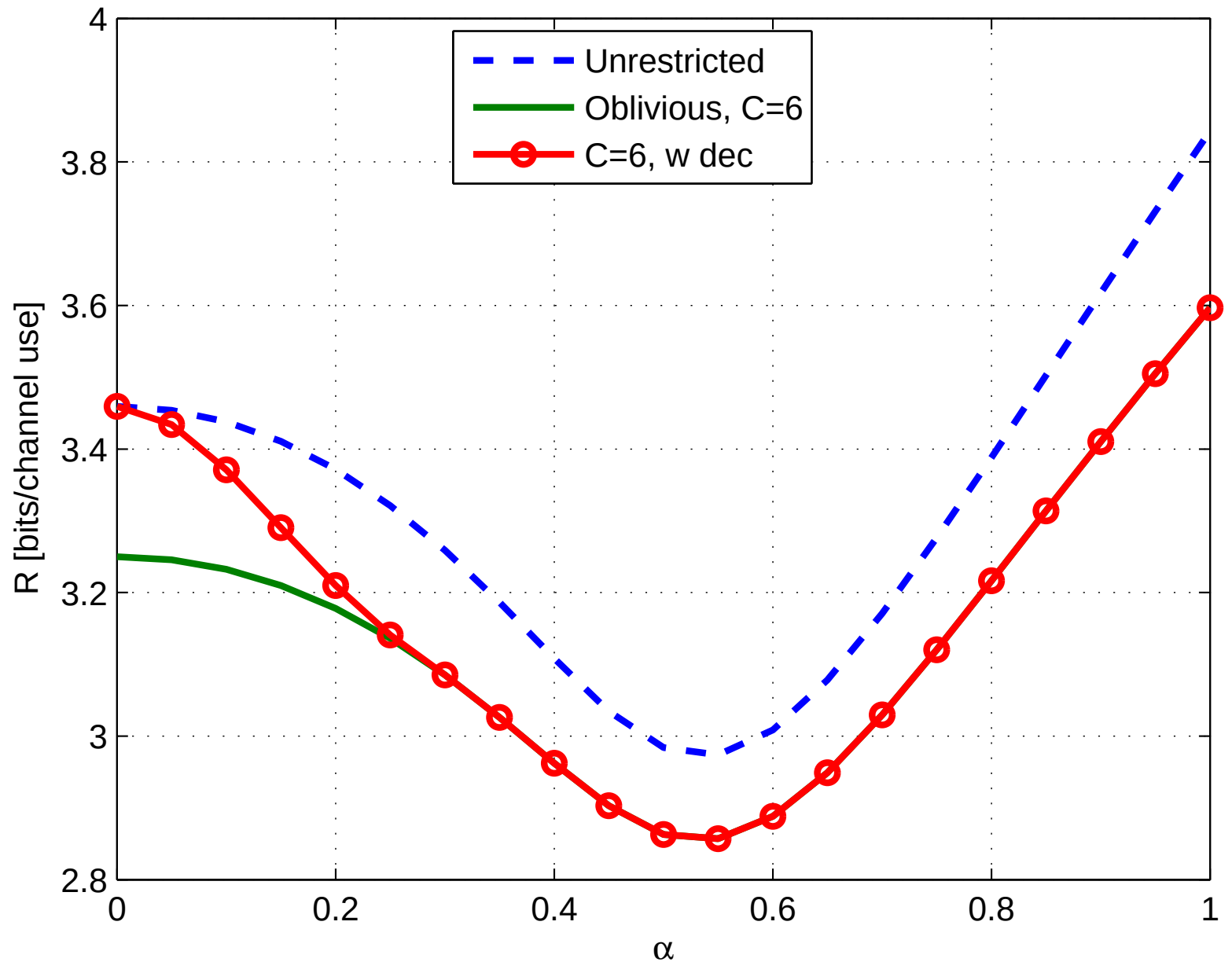

Fig. 3. The achievable rate for limited capacity $C = 6$ bits per channel use and oblivious processing (solid line) or combined with local decoding (circle marker), as compared with the unrestricted backhaul rate (dashed line), as a function of the interference level $\alpha$. The signal to noise ratio is $P = 10$ dB, and the channel is additive white Gaussian.

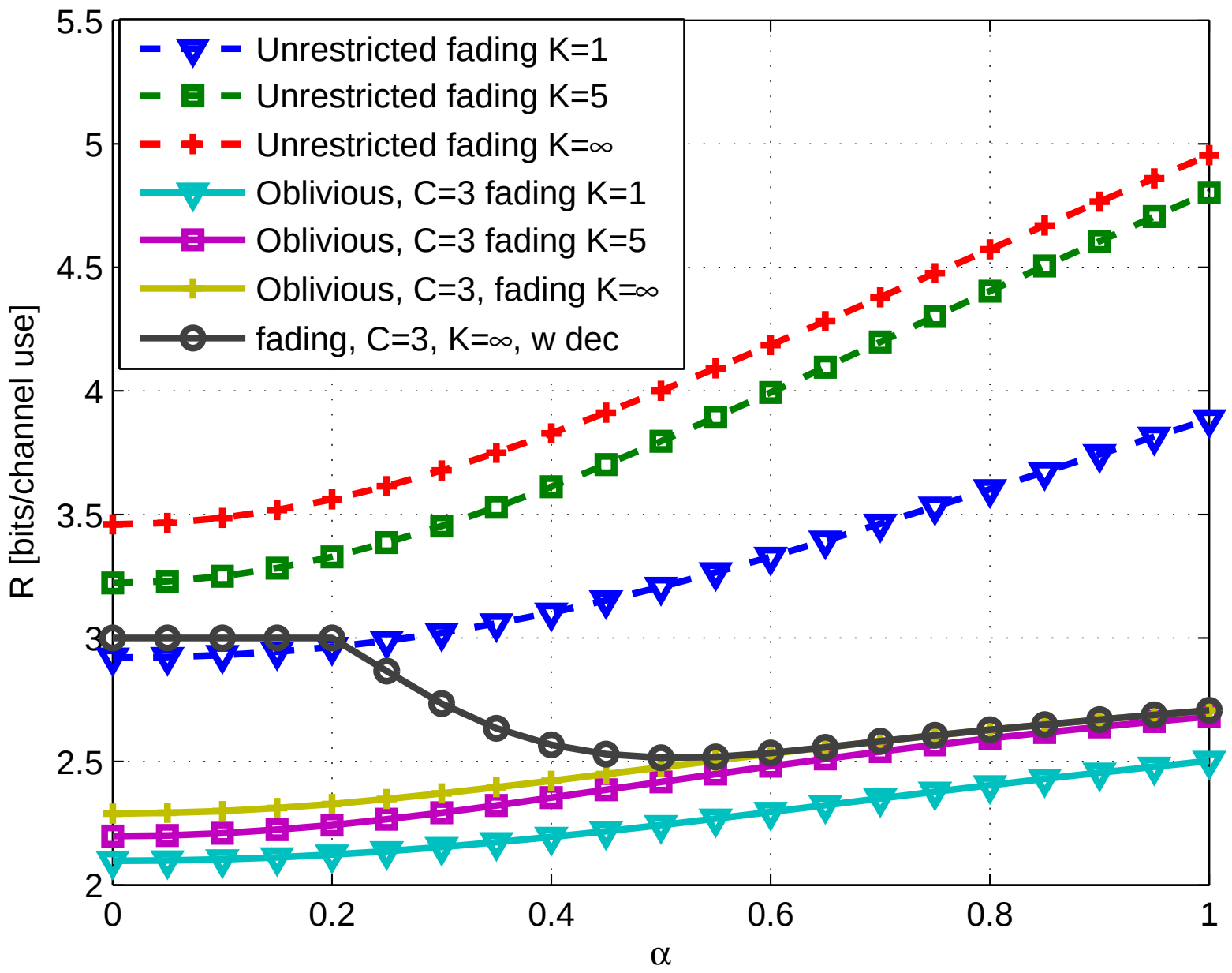

Fig. 4. The achievable rate for limited capacity $C = 3$ bits per channel use and oblivious processing (solid lines) or combined with local decoding (circle marker), as compared with the unrestricted backhaul rate (dashed lines), as a function of the interference level $\alpha$. The average per-cell sum signal to noise ratios is $P = 10$ dB, and the channel is Rayleigh flat fading. Three access protocols are plotted, with TDMA (triangular marker), with WB for five users (square) and with WB for infinitely many users (pluses).

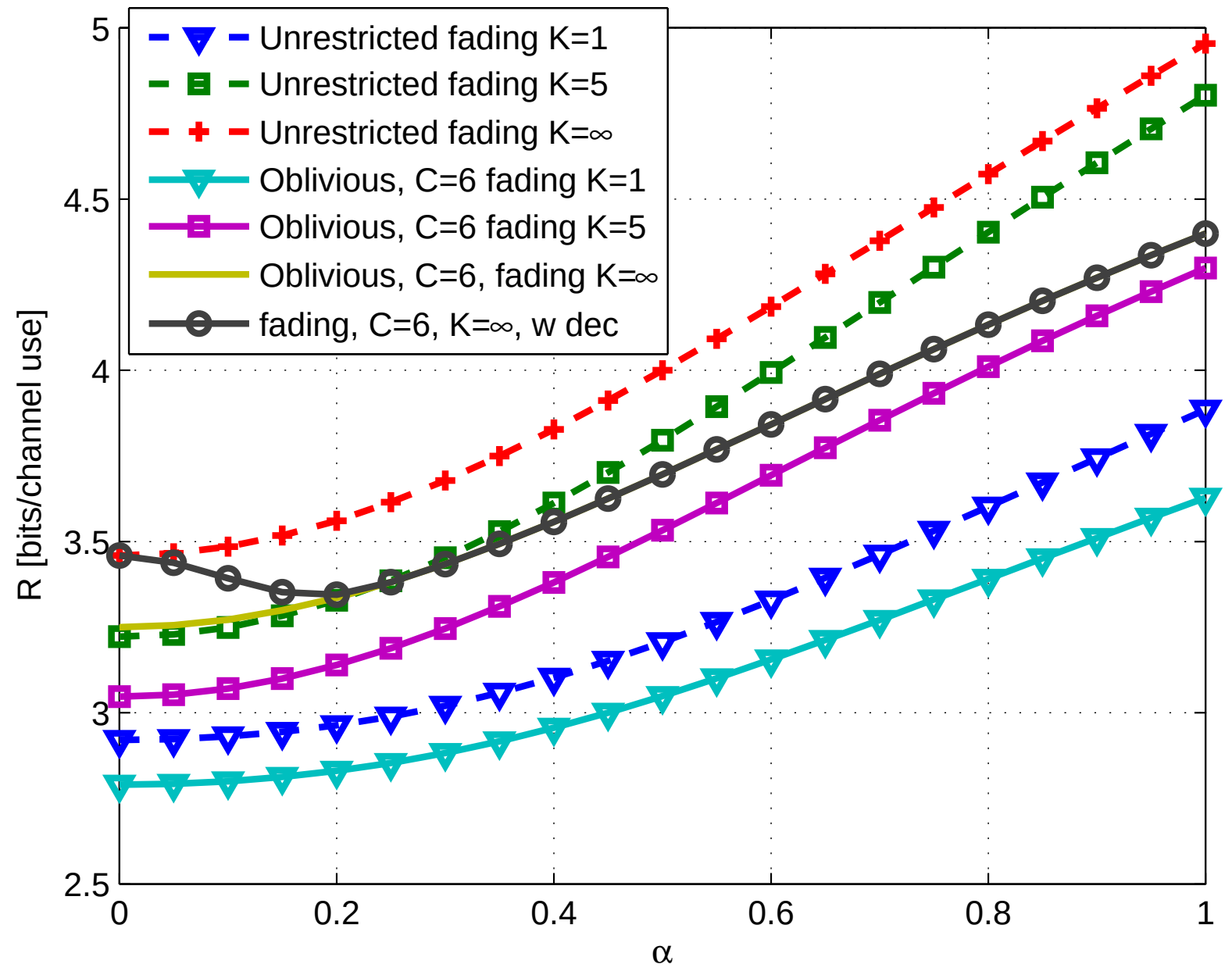

Fig. 5. The achievable rate for limited capacity $C = 6$ bits per channel use and oblivious processing (solid lines) or combined with local decoding (circle marker), as compared with the unrestricted backhaul rate (dashed lines), as a function of the interference level $\alpha$. The average per-cell sum signal to noise ratios is $P = 10$ dB, and the channel is Rayleigh flat fading. Three access protocols are plotted, with TDMA (triangular marker), with WB for five users (square) and with WB for infinitely many users (pluses).

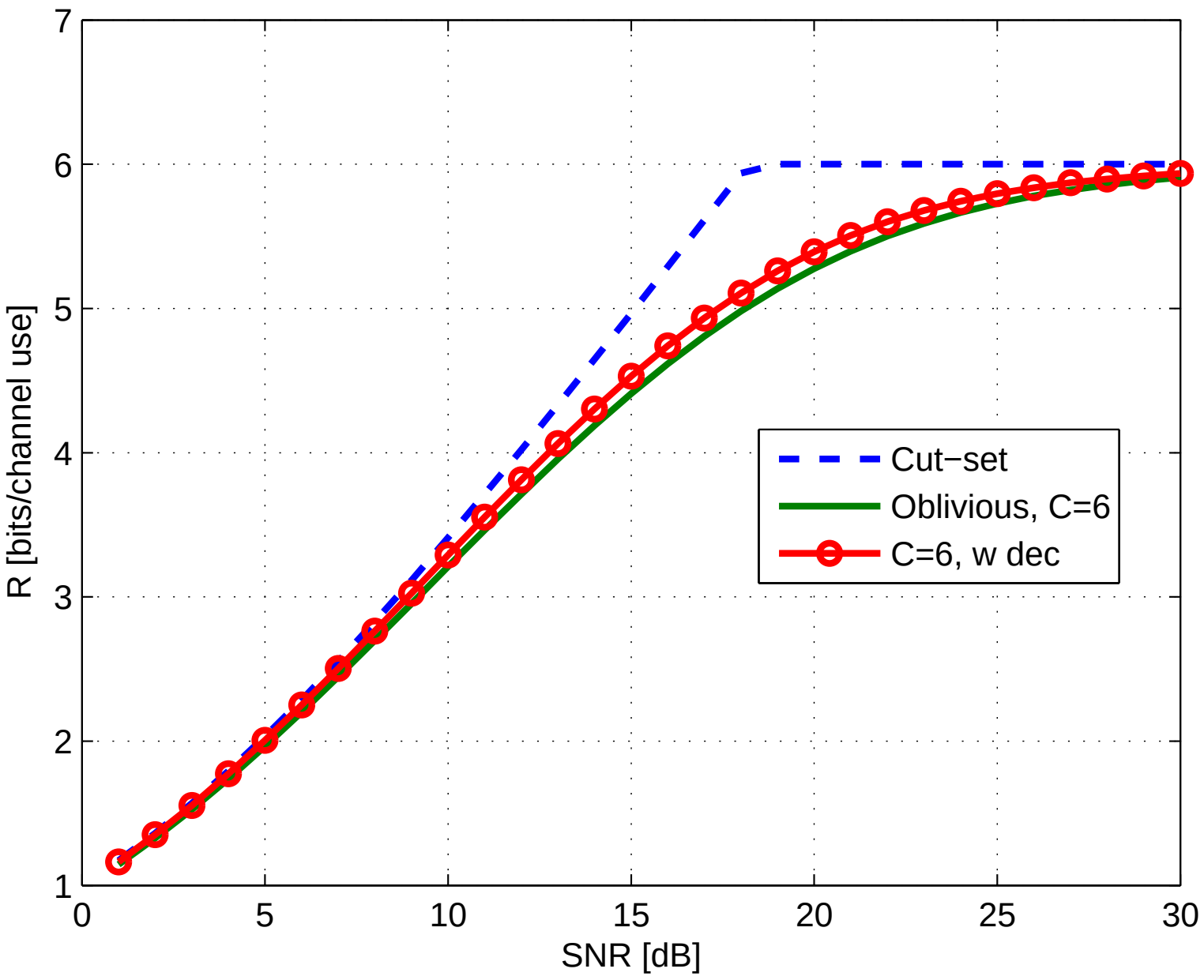

Fig. 6. The achievable rate for limited capacity $C = 6$ bits per channel use and oblivious processing (solid line) or combined with local decoding (circle marker), as compared with the upper bound (minimum of unrestricted backhaul rate and the backhaul rate - dashed line), as a function of the signal to noise ratio $P$ [dB], and the channel is additive white Gaussian. The interference level is $\alpha = 0.15$.

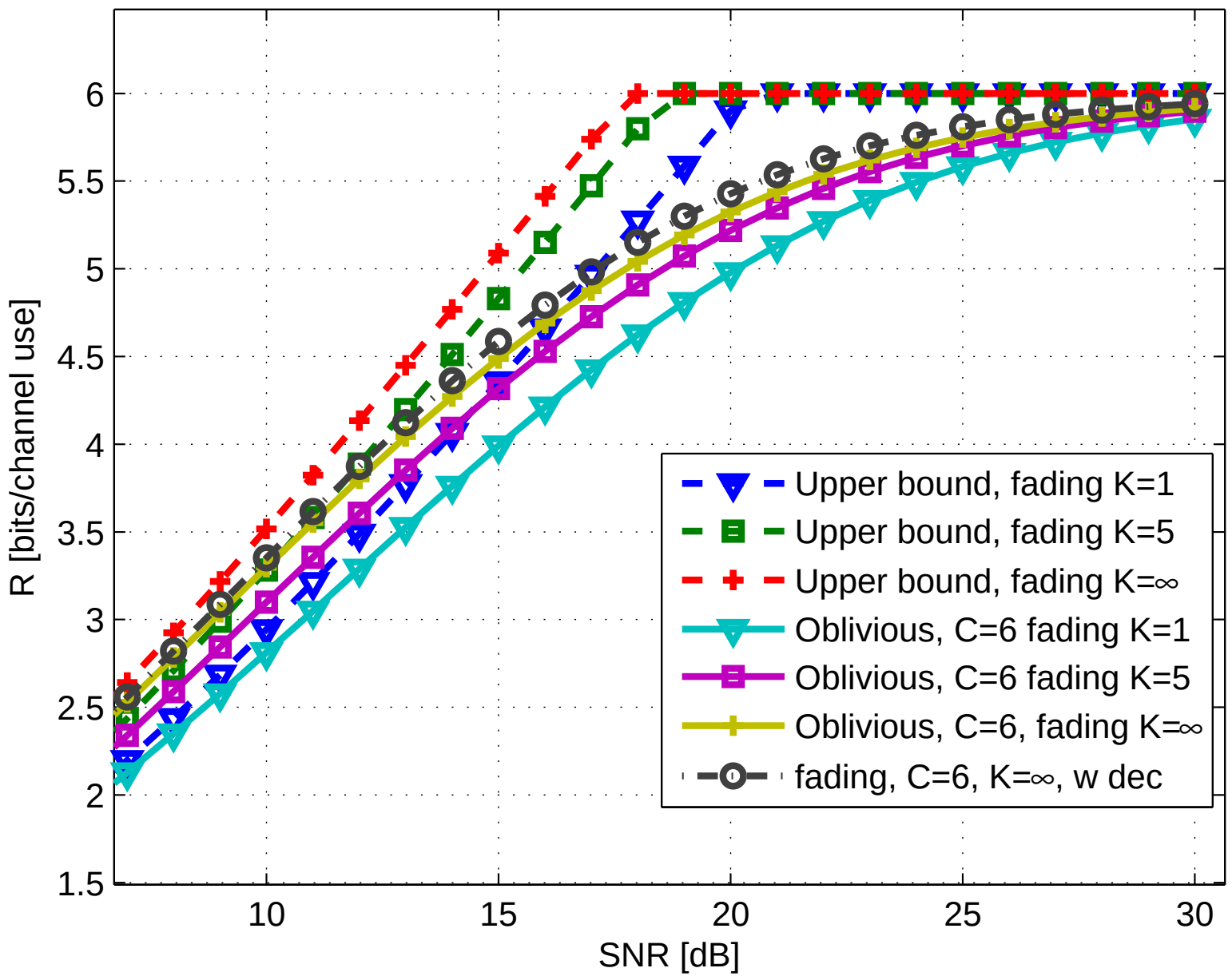

Fig. 7. The achievable rate for limited capacity $C = 6$ bits per channel use and oblivious processing (solid lines) or combined with local decoding (circle marker), as compared with the upper bound (minimum of unrestricted backhaul rate and the backhaul rate - dashed lines), as a function of the average signal to noise ratio $P$ [dB], when the channel is Rayleigh fading. The interference level is $\alpha = 0.15$. Three access protocols are plotted, with TDMA (triangular marker), with WB for five users (square) and with WB for infinitely many users (pluses).

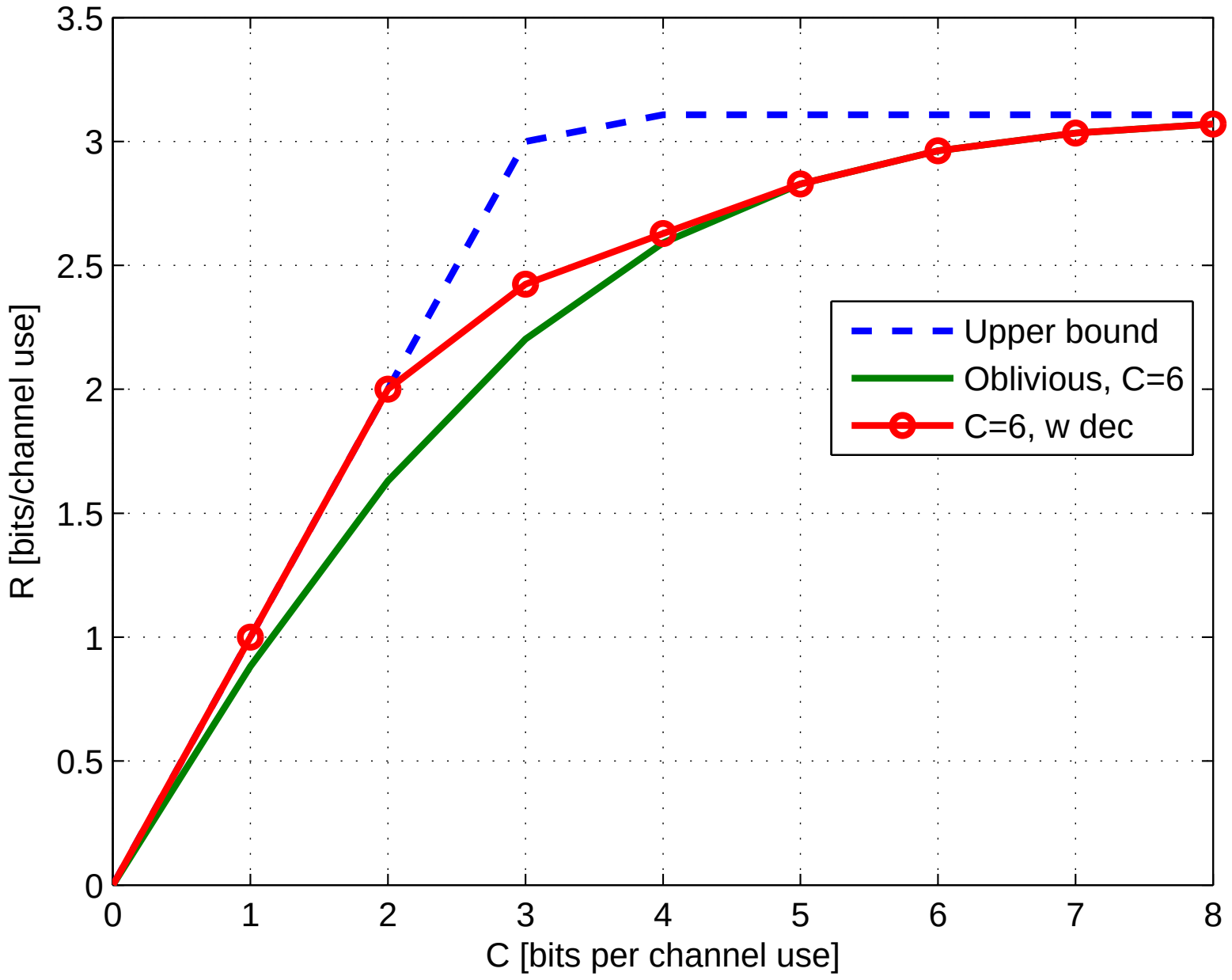

Fig. 8. The achievable rate for limited capacity $C$ bits per channel use and oblivious processing (solid line) or combined with local decoding (circle marker), as compared with the upper bound (minimum of unrestricted backhaul rate and the backhaul rate - dashed line), as a function of the backhaul capacity [bits per channel use]. The channel is additive white Gaussian with SNR $P = 10$ dB, and the interference level is $\alpha = 0.4$.

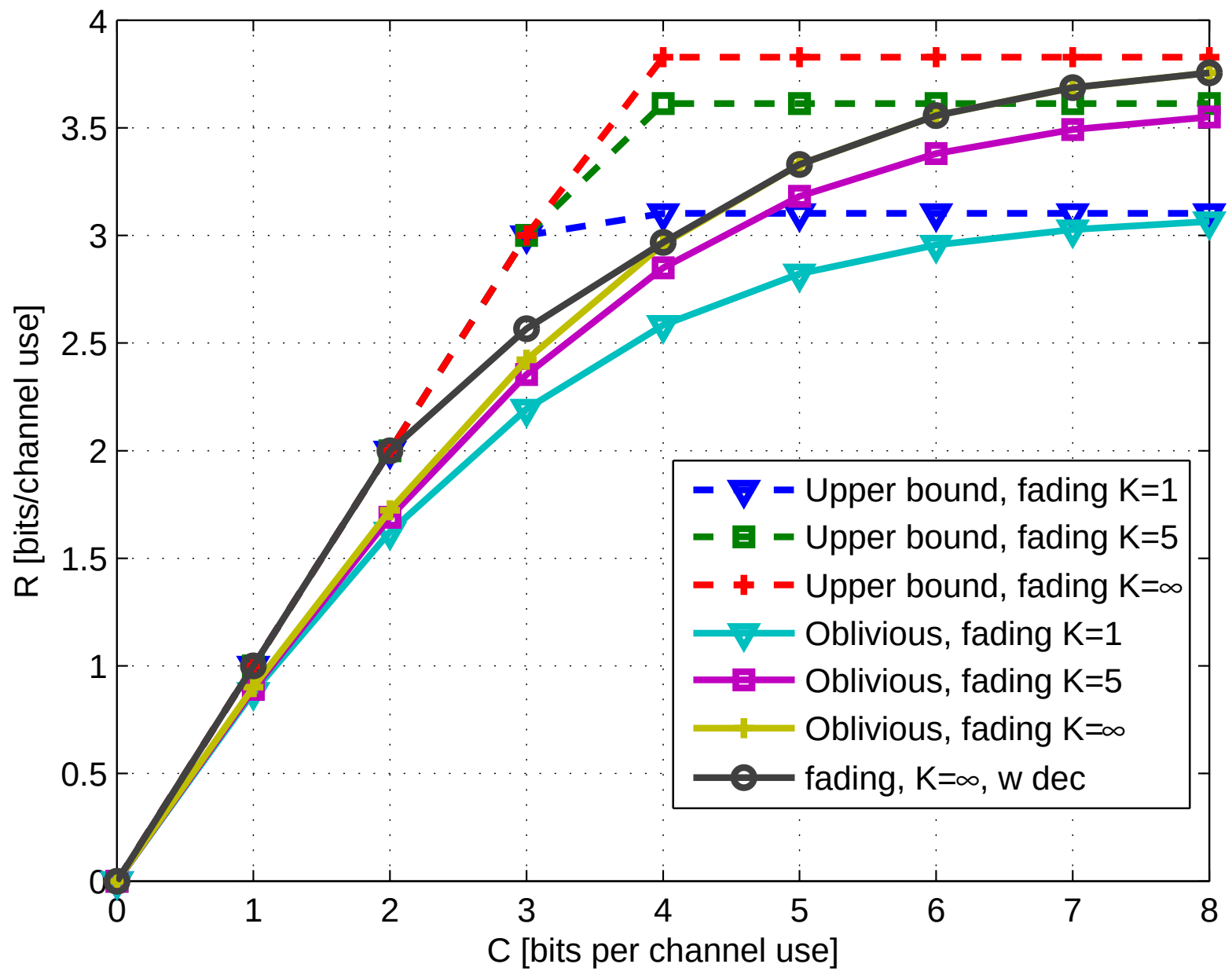

Fig. 9. The achievable rate for limited capacity $C$ bits per channel use and oblivious processing (solid lines) or combined with local decoding (circle marker), as compared with the upper bound (minimum of unrestricted backhaul rate and the backhaul rate - dashed lines), as a function of the backhaul capacity [bits per channel use]. The channel is Rayleigh fading with average SNR $P = 10$ dB, and the interference level $\alpha = 0.4$. Three access protocols are plotted, with TDMA (triangular marker), with WB for five users (square) and with WB for infinitely many users (pluses).

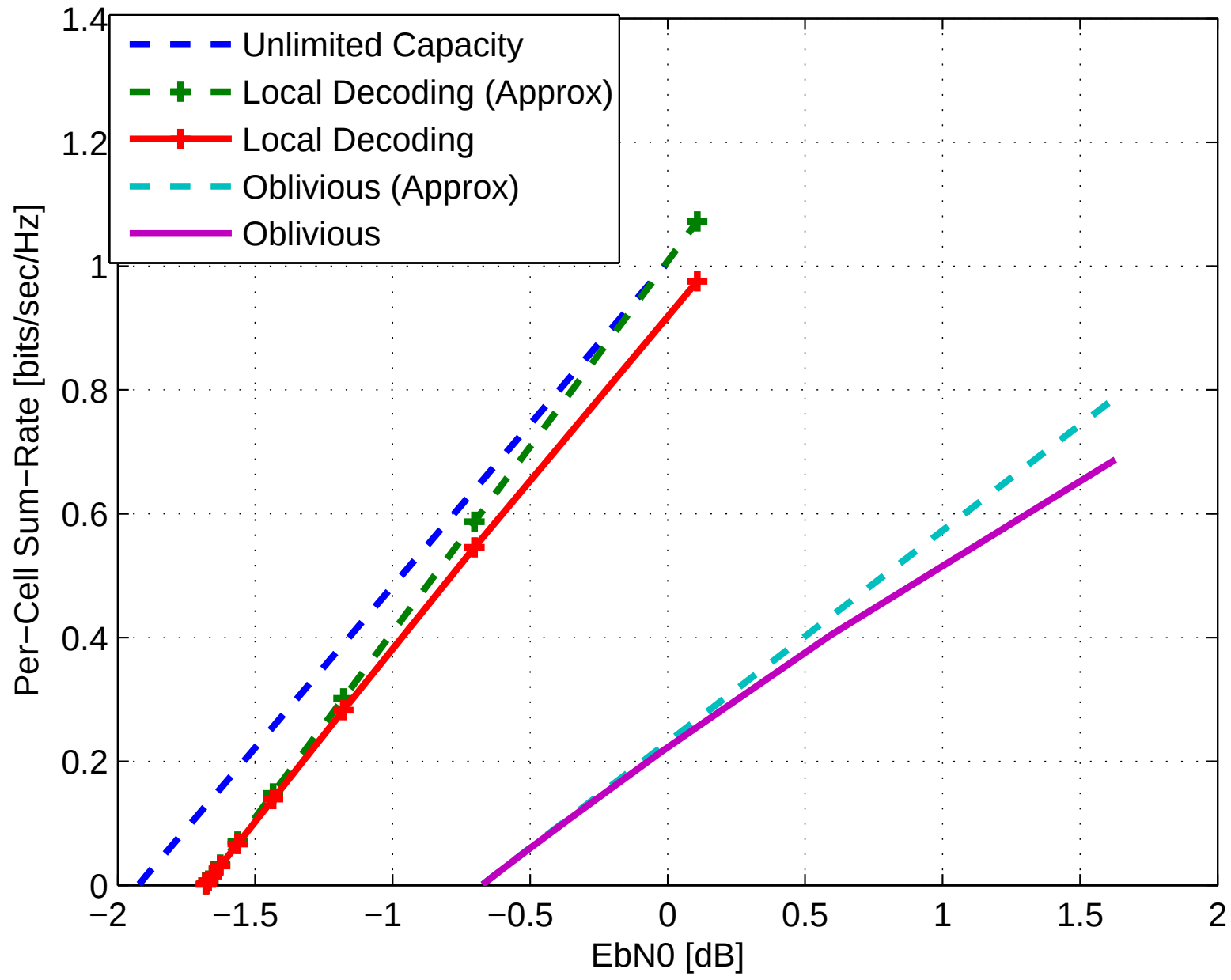

Fig. 10. Low-SNR region of the per-cell sum-rate, as a function of $\frac{E_b}{N_0}$ for oblivious processing (solid line) and local decoding (plus marker) along with the corresponding approximation (dashed) and the upper bound (upper dashed line). The channel is additive white Gaussian, with interference level $\alpha = 0.2$, and backhaul capacity of 2 [bits/channel].

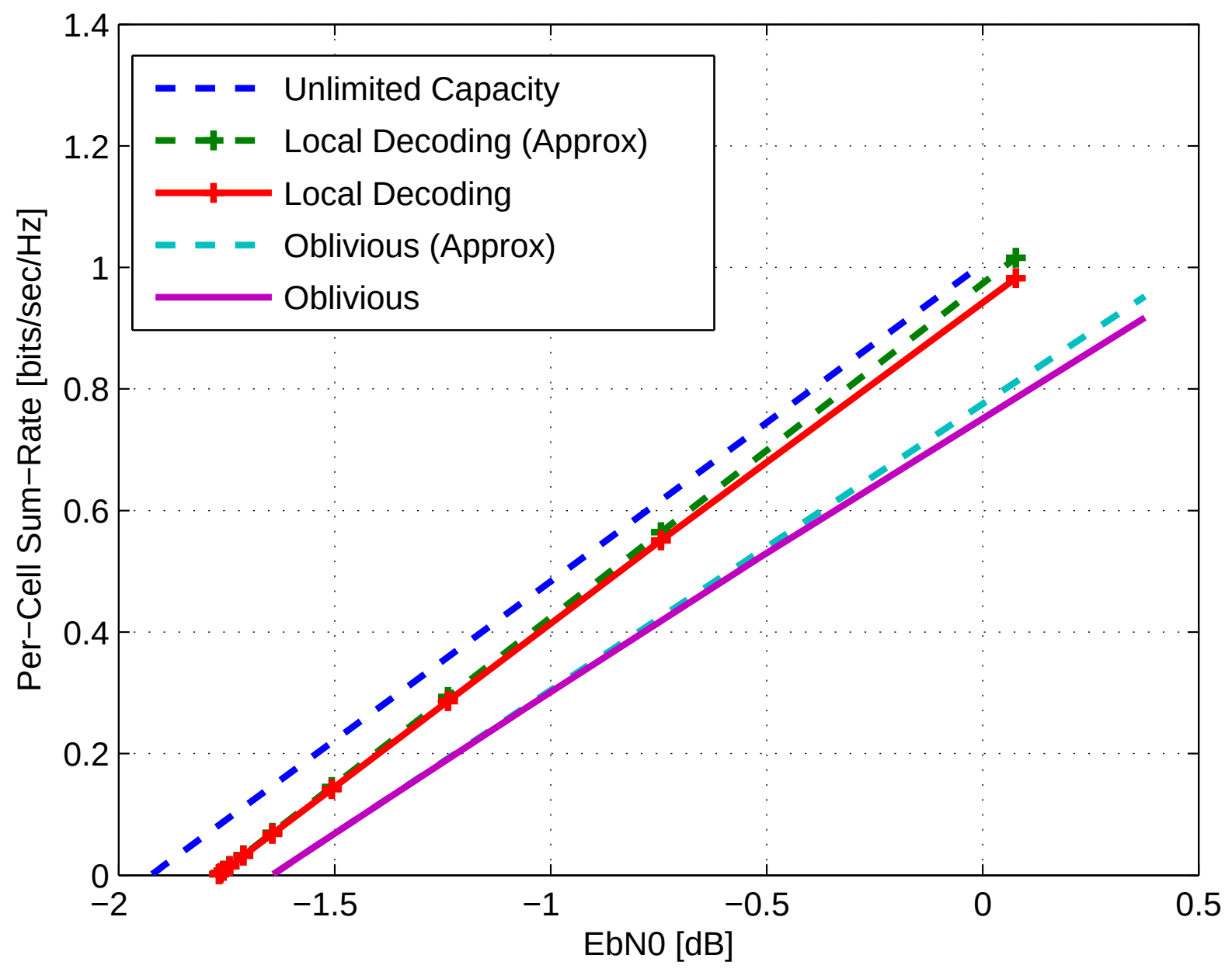

Fig. 11. Low-SNR region of the per-cell sum-rate, as a function of $\frac{E_b}{N_0}$ for oblivious processing (solid line) and local decoding (plus marker) along with the corresponding approximation (dashed) and the upper bound (upper dashed line). The channel is additive white Gaussian, with interference level $\alpha = 0.2$, and backhaul capacity of 4 [bits/channel].

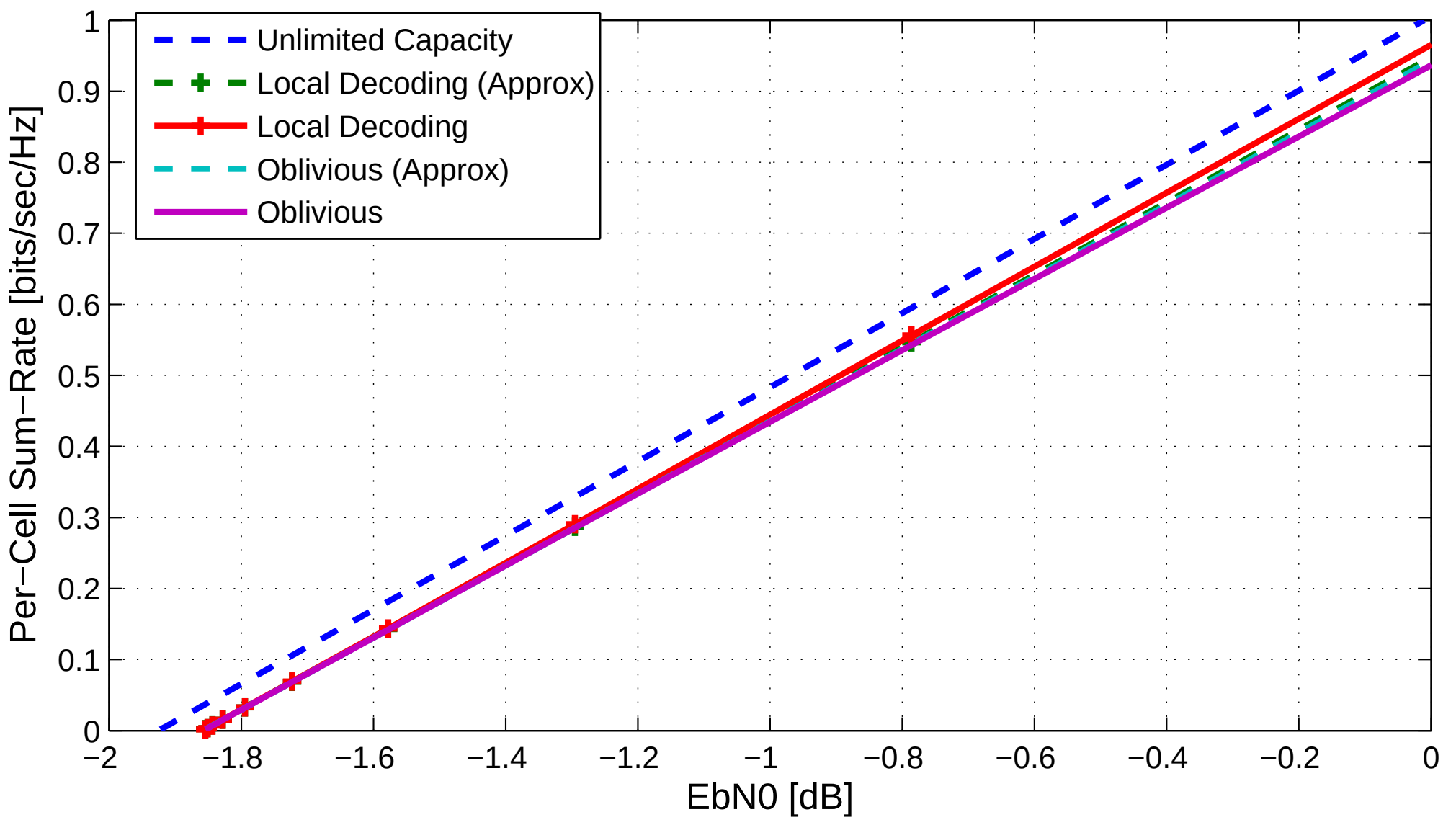

Fig. 12. Low-SNR region of the per-cell sum-rate, as a function of $\frac{E_b}{N_0}$ for oblivious processing (solid line) and local decoding (plus marker) along with the corresponding approximation (dashed) and the upper bound (upper dashed line). The channel is additive white Gaussian, with interference level $\alpha = 0.2$, and backhaul capacity of 6 [bits/channel].